\documentclass{emulateapj}
\usepackage{graphics,graphicx}
\usepackage{epstopdf}
\usepackage{amssymb, amsmath, amsthm}
\usepackage[normalem]{ulem}
\usepackage{float}
\usepackage{color}
\usepackage{xcolor}
\definecolor{ultramarine}{rgb}{0.01, 0.64, 0.86} 
\def\green#1 {{\textcolor{ultramarine}{#1}}\ }

\shortauthors{Lin et al.}

\shorttitle{Structure of IRDCs}
\shortauthors{Lin et al.}
 
\begin{document}

\title{Cloud structure of three Galactic infrared dark star-forming regions from combining ground and space based bolometric observations}

\author{Yuxin Lin\altaffilmark{1,2}}
\author{Hauyu Baobab Liu\altaffilmark{3}}
\author{James E. Dale\altaffilmark{4}}
\author{Di Li\altaffilmark{1,5}}
\author{Gemma Busquet\altaffilmark{6}}
\author{Zhi-Yu Zhang\altaffilmark{3,7}}
\author{Adam Ginsburg\altaffilmark{3}}
\author{Roberto Galv\'{a}n-Madrid\altaffilmark{8}}
\author{Attila Kov{\'a}cs\altaffilmark{9}}
\author{Eric Koch\altaffilmark{10}}
\author{Lei Qian\altaffilmark{1,5}}
\author{Ke Wang\altaffilmark{3}}
\author{Steve Longmore\altaffilmark{11}}
\author{Huei-Ru Chen\altaffilmark{12}}
\author{Daniel Walker\altaffilmark{11}}





 \affil{$^{1}$National Astronomical Observatories, Chinese Academy of Sciences}
 \affil{$^{2}$Max-Planck-Institut f\"{u}r Radioastronomie, D-53121 Bonn, Germany; \textcolor{blue}{ylin@mpifr-bonn.mpg.de}}
 \affil{$^{3}$European Southern Observatory (ESO), Karl-Schwarzschild-Str. 2, D-85748 Garching, Germany}
 \affil{$^{4}$Centre for Astrophysics Research, University of Hertfordshire, College Lane, Hatfield, AL10 9AB, UK}
 \affil{$^{5}$Key Laboratory of Radio Astronomy, Chinese Academy of Sciences}
 \affil{$^{6}$Institut de Ci\`encies de l'Espai (IEEC-CSIC), Campus UAB, Carrer de Can Magrans S/N, 08193, Barcelona, Catalunya, Spain}
  \affil{$^{7}$Institute for Astronomy, University of Edinburgh, Royal Observatory, Blackford Hill, Edinburgh EH9 3HJ, UK}
  \affil{$^{8}$Instituto de Radioastronom\'ia y Astrof\'isica, UNAM, Apdo. Postal 3-72 (Xangari), 58089 Morelia, Michoac\'an, M\'exico}
\affil{$^{9}$California Institute of Technology 301-17, 1200 E California Blvd, Pasadena, CA 91125, USA}
\affil{$^{10}$Department of Physics, University of Alberta, 4-181 CCIS, Edmonton, AB T6G 2E1, Canada}
 \affil{$^{11}$Astrophysics Research Institute, Liverpool John Moores University, 146 Brownlow Hill, Liverpool L3 5RF, UK}
 \affil{$^{12}$Institute of Astronomy and Department of Physics, National Tsing Hua University, Hsinchu, Taiwan}

\begin{abstract}
We have modified the iterative procedure introduced by Lin et al. (2016), to systematically combine the submm images taken from ground based (e.g., CSO, JCMT, APEX) and space (e.g., Herschel, Planck) telescopes.
We applied the updated procedure to observations of three well studied Infrared Dark Clouds (IRDCs): G11.11-0.12, G14.225-0.506 and G28.34+0.06, and then performed single-component, modified black-body fits to each pixel to derive $\sim$10$''$ resolution dust temperature and column density maps.
The derived column density maps show that these three IRDCs exhibit complex filamentary structures embedding with rich clumps/cores. 
We compared the column density probability distribution functions (N-PDFs) and two-point correlation (2PT) functions of the column density field between these IRDCs with several OB cluster-forming regions. 
Based on the observed correlation between the luminosity-to-mass ratio and the power-law index of the N-PDF, and complementary hydrodynamical simulations for a 10$^{4}$ $\rm M_{\odot}$ molecular cloud, we hypothesize that cloud evolution can be better characterized by the evolution of the (column) density distribution function and the relative power of dense structures as a function of spatial scales, rather than merely based on the presence of star-forming activity.
An important component of our approach is to provide a model-independent quantification of cloud evolution.
Based on the small analyzed sample, we propose four evolutionary stages, namely: {\it cloud integration, stellar assembly, cloud pre-dispersal and dispersed-cloud.} The initial {\it cloud integration} stage and the final  {\it dispersed cloud} stage may be distinguished from the two intermediate stages by a steeper than $-$4 power-law index of the N-PDF.
The {\it cloud integration} stage and the subsequent {\it stellar assembly} stage are further distinguished from each other by the larger luminosity-to-mass ratio ($>$40 $\rm L_{\odot}/M_{\odot}$) of the latter.
A future large survey of molecular clouds with high angular resolution may establish more precise evolutionary tracks in the parameter space  of N-PDF, 2PT function, and luminosity-to-mass ratio.
\end{abstract}


\section{Introduction}
Infrared dark clouds (IRDCs) are dense, cold molecular gas clouds that efficiently absorb the Galactic bright mid-infrared emission (Hennebelle et al. 2001; Simon et al. 2006; Peretto \& Fuller 2009). 
Due to the high molecular gas mass and low bolometric luminosity of these objects, they are considered candidate progenitors of massive stars or star clusters (Carey et al. 1998; Sridharan et al. 2005; Ragan et al. 2006). 
The density distribution in the IRDCs may reflect the dominant physical mechanisms during the formation of these clouds and can be used to gauge their subsequent gravitational contraction.
In particular, the widely resolved filamentary morphologies of the IRDCs indicate that they will likely undergo a hierarchical collapse (Inutsuka \& Miyama 1997; V{\'a}zquez-Semadeni et al. 2007, 2009).
The details of how IRDCs collapse and eventually evolve into luminous OB cluster-forming molecular clouds remain uncertain.
Not all IRDCs will evolve to form massive stars (Peretto \& Fuller 2009; Kauffmann \& Pillai 2010). 
How to distinguish those possible progenitors of high-mass stars or clusters from large sample of IRDCs quantitatively and qualitatively remains not elaborately resolved. 

\begin{table*}\footnotesize{ 
\caption{Source information}
\label{tab:source_info}
\hspace{3.5cm}
\begin{tabular}{lcccccccc}\hline\hline

Target Source&RA&DEC & Distance$^{\mbox{\tiny{a}}}$ & Mass$^{\mbox{\tiny{b}}}$ & Luminosity$^{\mbox{\tiny{b}}}$ \\
&(J2000)  &(J2000)  &(kpc)& $(\mathrm{M_{\odot}})$ & $(\mathrm{L_{\odot}})$\\

\hline 

  G11.11-0.12&19$^{\mbox{\scriptsize{h}}}$10$^{\mbox{\scriptsize{m}}}$13$^{\mbox{\scriptsize{s}}}$.00&09$^{\circ}$06$'$00$''$.0 & $3.6\pm0.72$ & 9.3$\times10^{4}$ & 6.0$\times10^{5}$\\     
 
G14.225-0.506&18$^{\mbox{\scriptsize{h}}}$47$^{\mbox{\scriptsize{m}}}$36$^{\mbox{\scriptsize{s}}}$.43&-01$^{\circ}$59$'$02$''$.5& ${1.98^{+0.13}_{-0.12}}^{\mbox{\tiny{c}}}$ & 1.9$\times10^{4}$ & 1.95$\times10^{5}$\\

G28.34+0.06 &18$^{\mbox{\scriptsize{h}}}$46$^{\mbox{\scriptsize{m}}}$02$^{\mbox{\scriptsize{s}}}$.08&-02$^{\circ}$43$'$00$''$.8 & $4.8\pm0.96$ & 2.8$\times10^{4}$ & 1.3$\times10^{6}$ \\

\hline                     
\end{tabular}
\vspace{0.1cm}
\par
\scriptsize{
\begin{itemize}
\item[$^{\mbox{\scriptsize{a}}}$] We tentatively quote a 20\% distance uncertainty for kinematic distances. \vspace{-0.15cm}
\item[$^{\mbox{\scriptsize{b}}}$] Total masses were summed from our derived column density maps (Figure \ref{fig:TN_G11_hersext}-\ref{fig:TN_G28}) above each source's measured column density threshold. Total bolometric luminosity is calculated by integrating from 0.1 \micron\ to 1 cm of the obtained SED for each pixel of column density above threshold, and adding all the values in each field. For a detailed procedure of how these quantities are calculated, we refer to Lin et al. (2016).
\item[$^{\mbox{\scriptsize{c}}}$] Distance measurement by parallaxes of methanol masers (Xu et al. 2011; Wu et al. 2014).  \vspace{-0.15cm}
\end{itemize}
}
}
\vspace{0.1cm}
\end{table*}
To pilot the systematic quantification for the similarity and the difference between the IRDCs and the more evolved OB cluster-forming regions, we have performed high angular resolution mapping observations of the 350 $\mu$m dust emission towards three very well studied IRDCs, namely G11.11-0.12, G14.225-0.506, and G28.34+0.06, using the Caltech Submillimeter Observatory (CSO) Submillimetre High Angular Resolution Camera II (SHARC2).
We also performed deep 870 $\mu$m mapping observations on G14.225-0.506, using the Large Apex BOlometer CAmera (APEX-LABOCA; Siringo et al. 2009).
Table \ref{tab:source_info} summarizes the basic properties of the selected target sources.
References for these observed IRDCs can be found, for G11.11-0.12 in Pillai et al. (2006), Henning et al. (2011), Wang et al. (2014), and Pillai et al. (2015); for G28.34+0.06 in Wang et al. (2008), Zhang et al. (2009), Chen et al. (2011), Wang et al. (2011, 2012),  Bulter et al. (2014), and Zhang et al. (2015); and for G14.225-0506 in Busquet et al. (2013), Busquet et al. (2016), Santos et al. (2016).

\begin{figure*}
\hspace{-0.3cm}
\vspace{-0.1cm}
\includegraphics[width=22cm]{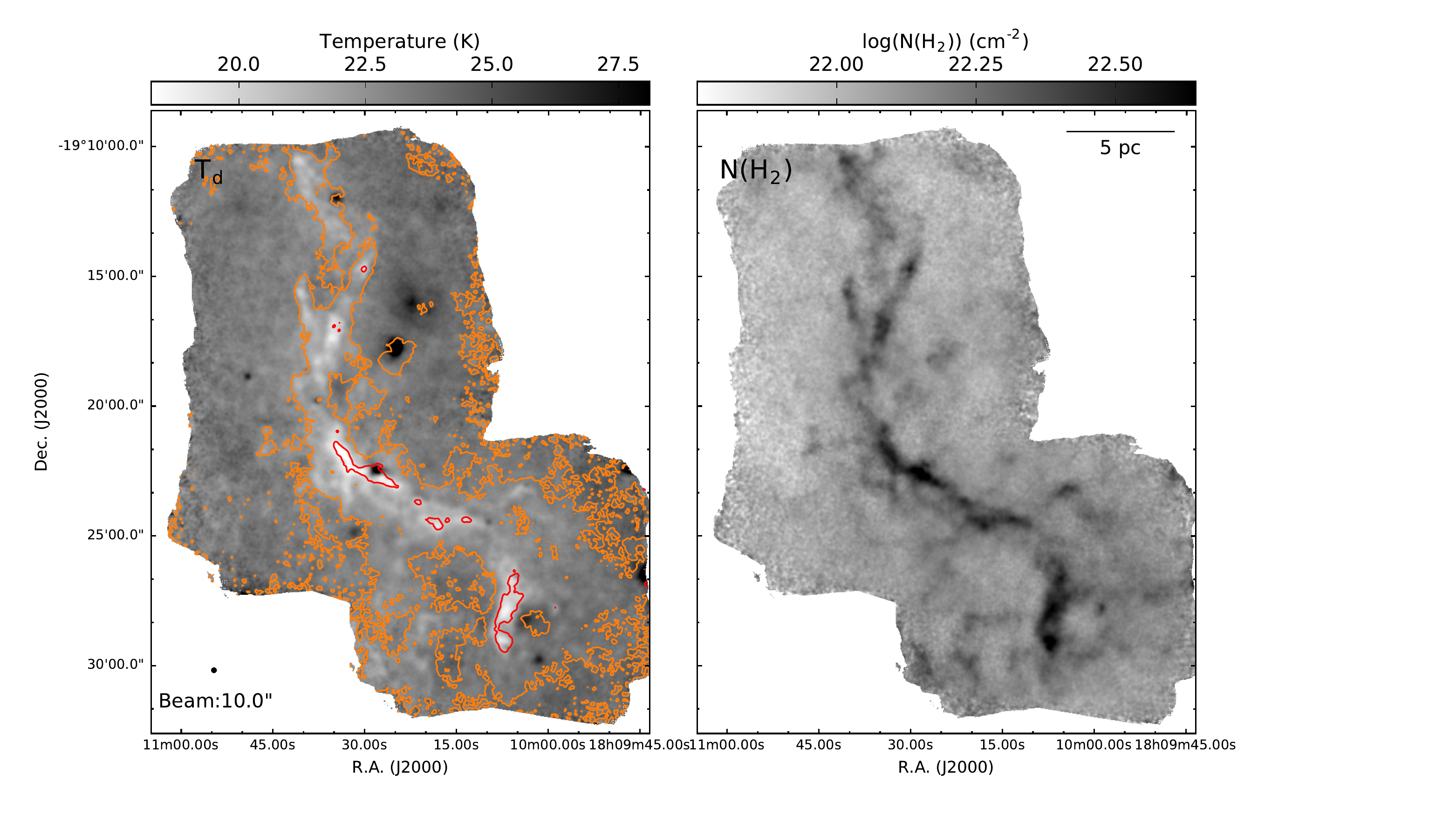}
\vspace{-0.6cm}
\caption{
Dust temperature and column density maps of G11.11-0.12, derived based on fitting modified black-body spectra iteratively to the Herschel-PACS 70/160 \micron\ , Herschel-SPIRE 250 \micron\ , and the combined 350/850 \micron\ images. The detailed procedures can be found in Section \ref{subsection:procedure} and Appendix A. Two contours are indicated in temperature map of column density levels $\rm log_{10}(N(H_{2}))$ of 22.14, 22.47. See also Figure \ref{fig:NPDF} for an explanation of how these contours probe the overall column density distribution.
}
\label{fig:TN_G11_hersext}
\vspace{0.3cm}
\end{figure*}

To obtain a finest possible angular resolution, and to precisely constrain physical properties on all angular scales, we build on procedures suggested by Liu et al. (2015) and Lin et al. (2016) to combine the 350 $\mu$m images and other ground based (JCMT-SCUBA2 and APEX-LABOCA, see Section \ref{section:obs}) 850/870 $\mu$m images with space telescopes observations, and then derive the dust temperature and dust/gas column density maps with $\sim$10$''$ angular resolution.
In addition, we derived the statistical measures of the cloud structures for these IRDCs, and then compare with the same analysis for seven very luminous ($\rm L_{bol}$$>$10$^{6}$ $\rm L_{\odot}$) OB cluster-forming molecular clouds published in Lin et al. (2016).
Finally, we performed numerical hydrodynamics simulations to demonstrate how the cloud morphology and the derived statistical quantities evolve with time.
The observations and data analysis procedures are outlined in Section \ref{section:obs}. 
Results are provided in Section \ref{section:results}.
A comparison among the observed star-forming regions and the comparison with numerical hydrodynamics simulations, are given in Section \ref{section:discussion}. 
The main conclusion and ending remarks are listed in Section \ref{section:conclusion}.


\begin{figure*}
\hspace{-0.3cm}
\vspace{-0.1cm}
\includegraphics[width=22cm]{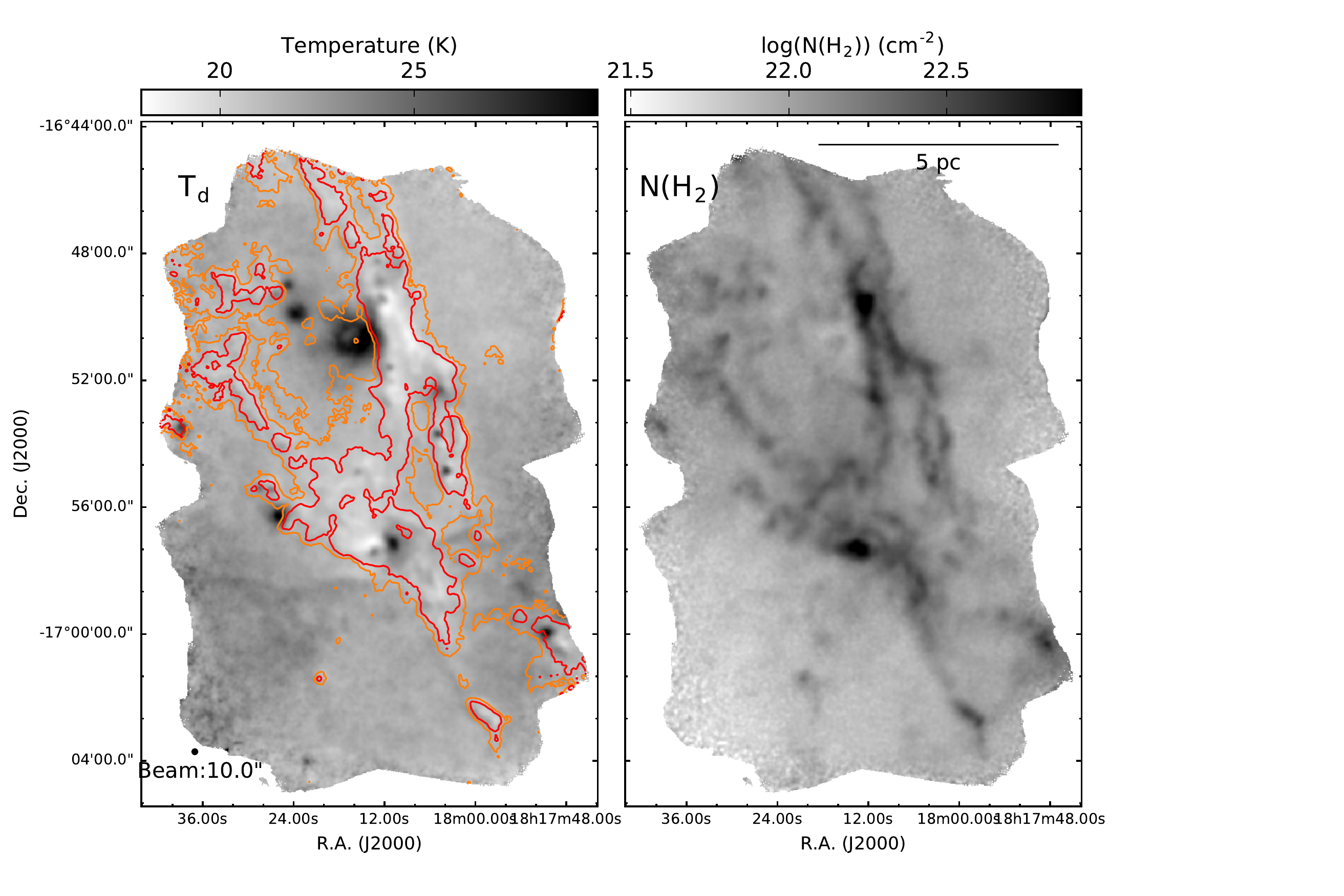}
\vspace{-0.6cm}
\caption{Same as Figure \ref{fig:TN_G11_hersext}, but for target source G14.225-0.506. Two contours are indicated in temperature map of column density levels $\rm log_{10}(N(H_{2}))$ of 22.09, 22.21.
}

\label{fig:TN_G14}
\vspace{0.3cm}
\end{figure*}

\section{Observations}\label{section:obs}
We introduce our CSO SHARC2 350 $\mu$m observations and data reduction in Section \ref{subsection:sharc2}.
The APEX-LABOCA observations and data calibrations are introduced in Section \ref{subsection:laboca}.
Section \ref{subsection:archive} outlines the archival data we included for the SED analysis.
The procedures for producing the final images and SED fitting are given in Section \ref{subsection:procedure}.

\begin{figure*}
\hspace{-0.3cm}
\vspace{-0.1cm}
\includegraphics[width=22cm]{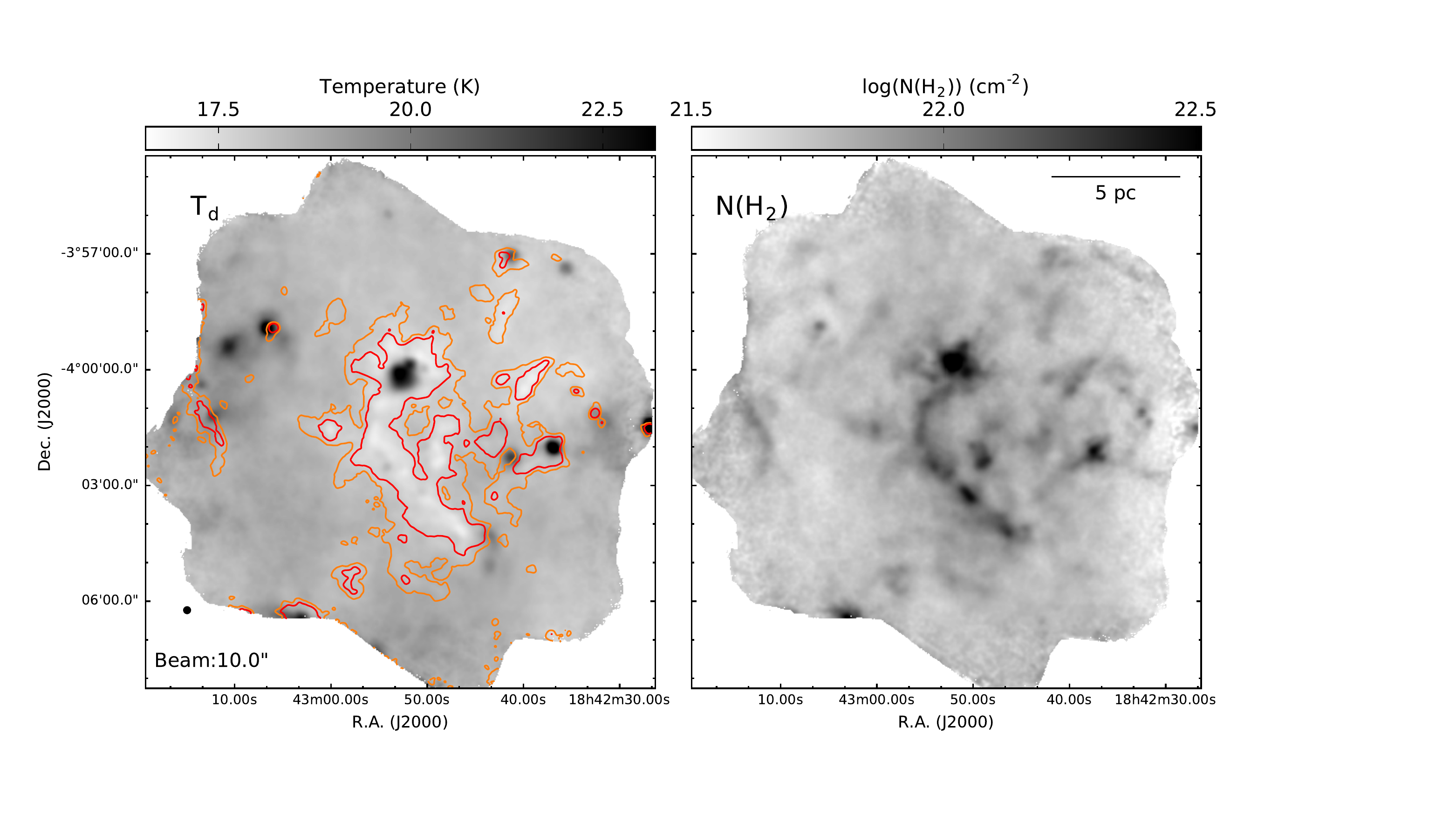}
\vspace{-1.4cm}
\caption{Same as Figure \ref{fig:TN_G11_hersext}, but for target source G28.34+0.06. Two contours are indicated in temperature map of column density levels $\rm log_{10}(N(H_{2}))$ of 21.83, 21.91.
}
\label{fig:TN_G28}
\vspace{0.3cm}
\end{figure*}

\subsection{CSO-SHARC2 observations}\label{subsection:sharc2}
High angular resolution, ground based continuum observations at 350 \micron\ towards the three IRDCs G11.11-0.12, G28.34+0.06 and G14.225-0.506 were carried out using the SHARC2 bolometer array, installed on the CSO Telescope (PI: H. B. Liu).
The array consists of 12$\times$32 pixels \footnote{Approximately 85\% of these pixels work well according to the online documentation http://www.astro.caltech.edu/$\sim$sharc/}.
The simultaneous field of view (FOV) provided by this array is 2$'$.59$\times$0$'$.97, and the diffraction limited beam size is $\sim$8$''$.8.
The data of G11.11-0.12 were acquired on March 28th 2014 ($\tau_{\mbox{\scriptsize{225 GHz}}}$$\sim$0.06), with an on-source exposure time of 90 minutes. 
G14.225-0.506 and G28.34+0.06 were observed on March 27th 2014 ($\tau_{\mbox{\scriptsize{225 GHz}}}$$\sim$0.05), with 80 and 50 minutes of on-source exposure time, respectively.
 The telescope pointing and focusing were checked every 1.5-2.5 hours.
Mars was observed for absolute flux calibration.
We used the standard 10$'$$\times$10$'$ on-the-fly (OTF) box scanning pattern, and the scanning center for each source is listed in Table 1.
Basic data calibration was carried out using the CRUSH software package (Kov\'{a}cs 2008).
We used the {\tt -faint} option of the CRUSH software package during data reduction, which optimized the reconstruction of the faint and compact sources with the cost of the more aggressive filtering of extended emission.
Nevertheless, the extended emission components will ultimately be complemented by the observations of space telescopes (see Section \ref{subsection:procedure} for more details).
The final calibrated map was smoothed with a Gaussian kernel with 2/3 beam FWHM ({\tt -faint} option) to an angular resolution of 9$''$.6 for an optimized sensitivity and source reconstruction, and the rms noise levels we measured from the approximately emission-free areas of 56, 53, 60 mJy\,beam$^{-1}$ for G11.11-0.12, G28.34+0.06 and G14.225-0.506, respectively.

\subsection{APEX-LABOCA observations}\label{subsection:laboca}
The 870~$\mu$m submillimeter continuum observations toward G14.225-0.506 were conducted (PI: G. Busquet) using LABOCA bolometer array, installed on the Atacama Pathfinder EXperiment (APEX)\footnote{This publication is based on data acquired with the Atacama Pathfinder Experiment (APEX). APEX is a collaboration between the Max-Planck-Institut fur Radioastronomie, the European Southern Observatory, and the Onsala Space Observatory}.
The array consists of 259 channels arranged in 9 concentric hexagons around the central channel. The field of view of the array is $11\farcm4$, and the angular resolution of each beam is $18\farcs6\pm1''$.
The observations were carried out on August 24th and 31st 2008 under 
weather conditions with zenith opacity values ranged from 0.15 to 0.24 at 870~$\mu$m. 
The observations were performed using a spiral raster mode mapping, providing a fully sampled and homogeneously covered map in an area of $15'\times15'$. 

Calibration was done using observations of Mars as well as secondary calibrators. The absolute flux uncertainty is estimated to be $\sim$8~\%. Pointing was checked every hour, finding a rms pointing accuracy of $2''$, and focus settings were performed once per night and during the sunset. The data was reduced using \texttt{MiniCRUSH} software package (see K\'ovacs 2008). The data reduction process included flat-fielding, opacity correction, calibration, correlated noise removal, and de-spiking. Further details of the data acquisition and data reduction are described in Busquet et al. (2016).

\subsection{Archival Herschel, Planck, JCMT data}\label{subsection:archive}
We retrieved the available night observations\footnote{http://www.eaobservatory.org/jcmt/science/archive/guide/} of James Clerk Maxwell Telescope (JCMT)\footnote{The James Clerk Maxwell Telescope is operated by the East Asian Observatory on behalf of The National Astronomical Observatory of Japan, Academia Sinica Institute of Astronomy and Astrophysics, the Korea Astronomy and Space Science Institute, the National Astronomical Observatories of China and the Chinese Academy of Sciences (Grant No. XDB09000000), with additional funding support from the Science and Technology Facilities Council of the United Kingdom and participating universities in the United Kingdom and Canada. The James Clerk Maxwell Telescope has historically been operated by the Joint Astronomy Centre on behalf of the Science and Technology Facilities Council of the United Kingdom, the National Research Council of Canada and the Netherlands Organization for Scientific Research. Additional funds for the construction of SCUBA-2 were provided by the Canada Foundation for Innovation.}  Submillimetre Common-User Bolometer Array 2 (SCUBA2)
 (Dempsey et al. 2013;
\ Chapin et al. 2013;
\ Holland et al. 2013
) at 850 \micron\ from the online data archive. (Program ID is M11BEC30 for G11.11-0.12 and G28.34+0.06). 
Ancillary data also includes level 2.5 and level 3 processed, archival {\it Herschel}\footnote{Herschel is an ESA space observatory with science instruments provided by European-led Principal Investigator consortia and with important participation from NASA.} images, which were taken by the Herschel Infrared Galactic Plane (Hi-GAL) survey (Molinari et al. 2010) 
at 70/160 \micron\ using the PACS instrument (Poglitsch et al. 2010) and at 250/350/500 \micron\ using the SPIRE instrument (Griffin et al. 2010). Observation IDs for G14.225-0.506 are 1342218997, 1342219000, for G11.11-0.12 are 1342204952, 1342218965 and for G28.34+0.06 is 1342218694. 
Planck/High Frequency Instrument (HFI) 353 GHz images are also used. 
For our combination purpose, we use the Herschel SPIRE extended emission products from the data base. 
The Planck 353 GHz images which are in units of $\rm K_{CMB}$ are converted to Jy\,beam$^{-1}$ (Zacchei et al. 2011; Planck HFI Core Team 2011a, 2011b).

\subsection{Image combination, and derivations of dust column density and temperature}\label{subsection:procedure}
Our procedure to combine images taken from ground based and space telescope observations in the Fourier domain, and then iteratively derive the high angular resolution dust temperature and dust/gas column density images, is similar to what was introduced in Lin et al. (2016).
Since the observed IRDCs for the present paper have relatively low temperatures, it is particularly important to precisely determine the long wavelength part of the spectrum. 
Therefore, we additionally included the following steps to improve the quality of the combined 850 $\mu$m image before making our final modified black-body spectral energy distribution (SED) fits.

We extrapolated the 850 $\mu$m flux from SED fits to {\it Herschel} PACS 160 $\mu$m and SPIRE 250/350/500 $\mu$m. 
In these fits, we fixed the dust emissivity index $\beta$ to 1.8, which is the mean value measured for Galactic disk (Planck Collaboration 2011, 2014). 
The {\it Herschel}-extrapolated  850 $\mu$m image has the same resolution as SPIRE 500 $\mu$m,  $\sim$37$''$.  
Then we proceeded to combine the above mentioned 850 $\mu$m image with {\it Planck} 353 GHz, and then further combined the result with the SCUBA2 850 $\mu$m image.
In this way, the $>$5$'$ scale and the $<$3$'$-4$'$ scale structures are dominantly constrained by the {\it Planck} and the SCUBA2 images, respectively.
The role of the {\it Herschel}-extrapolated  850 $\mu$m image is to complement the small range of spatial scales which is poorly sampled by {\it Planck} and SCUBA2.

\begin{figure}[h]
\begin{tabular}{p{10cm}}
\includegraphics[width=9.5cm]{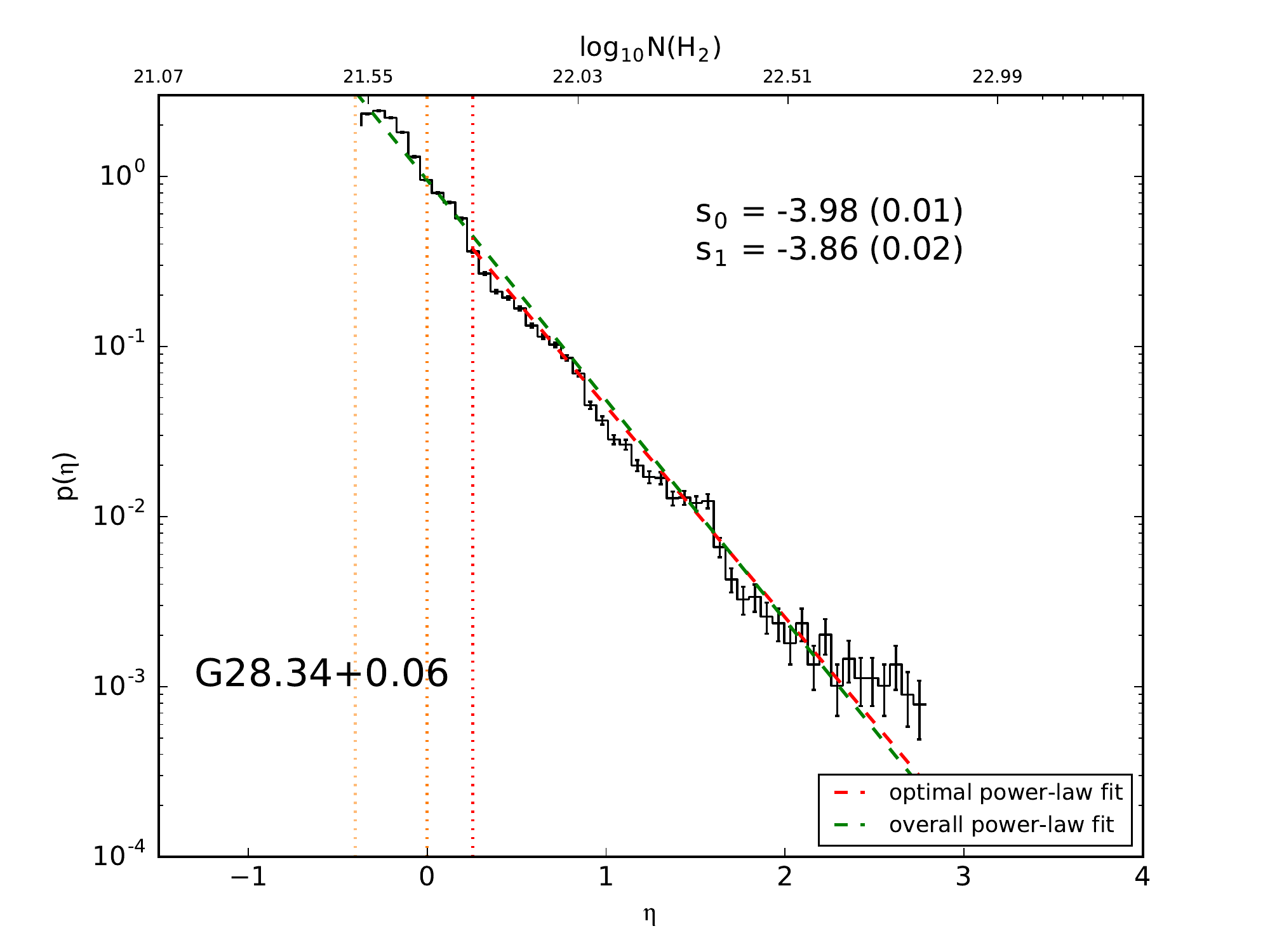}\\
\includegraphics[width=9.5cm]{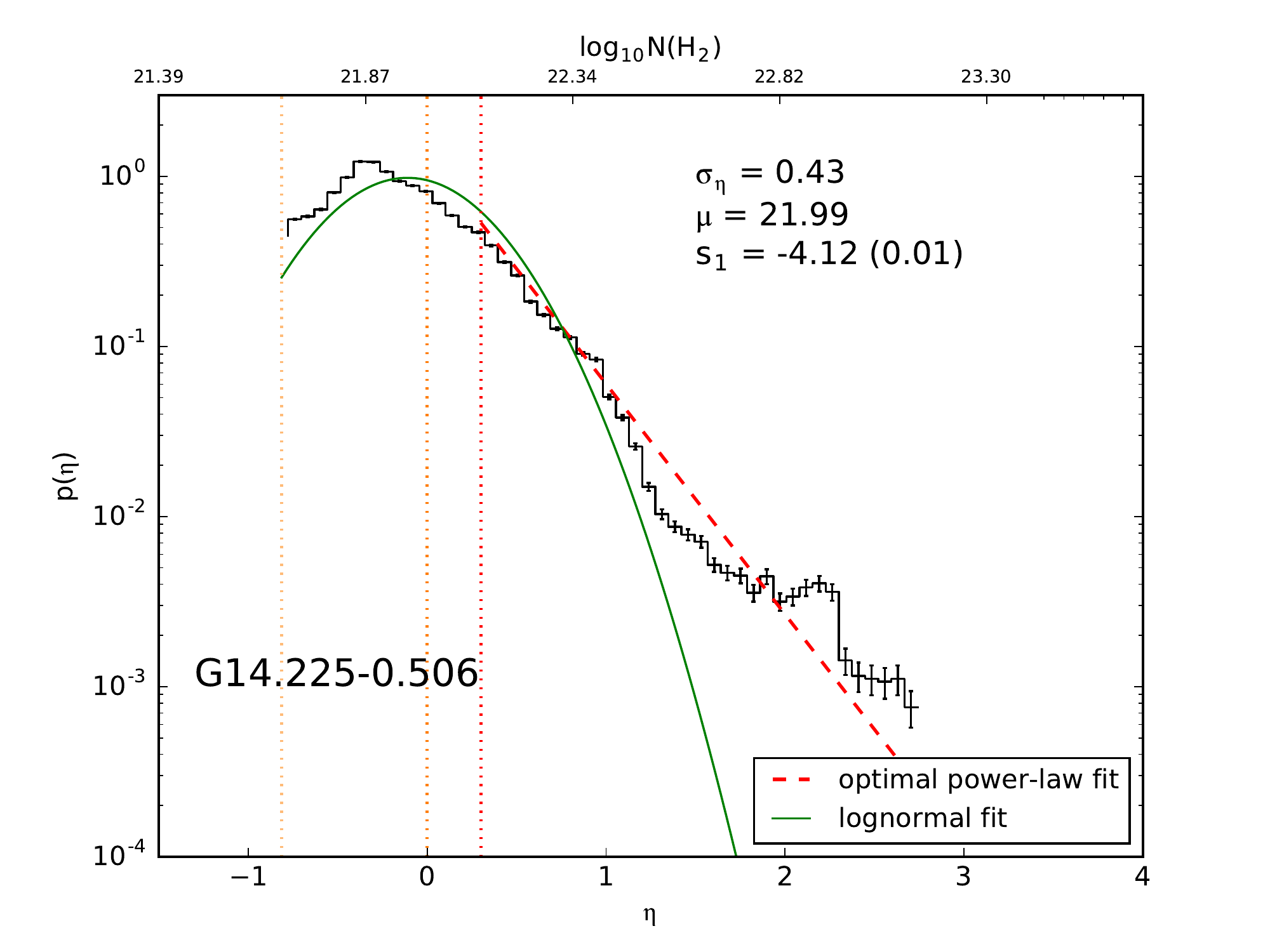}\\
\includegraphics[width=9.5cm]{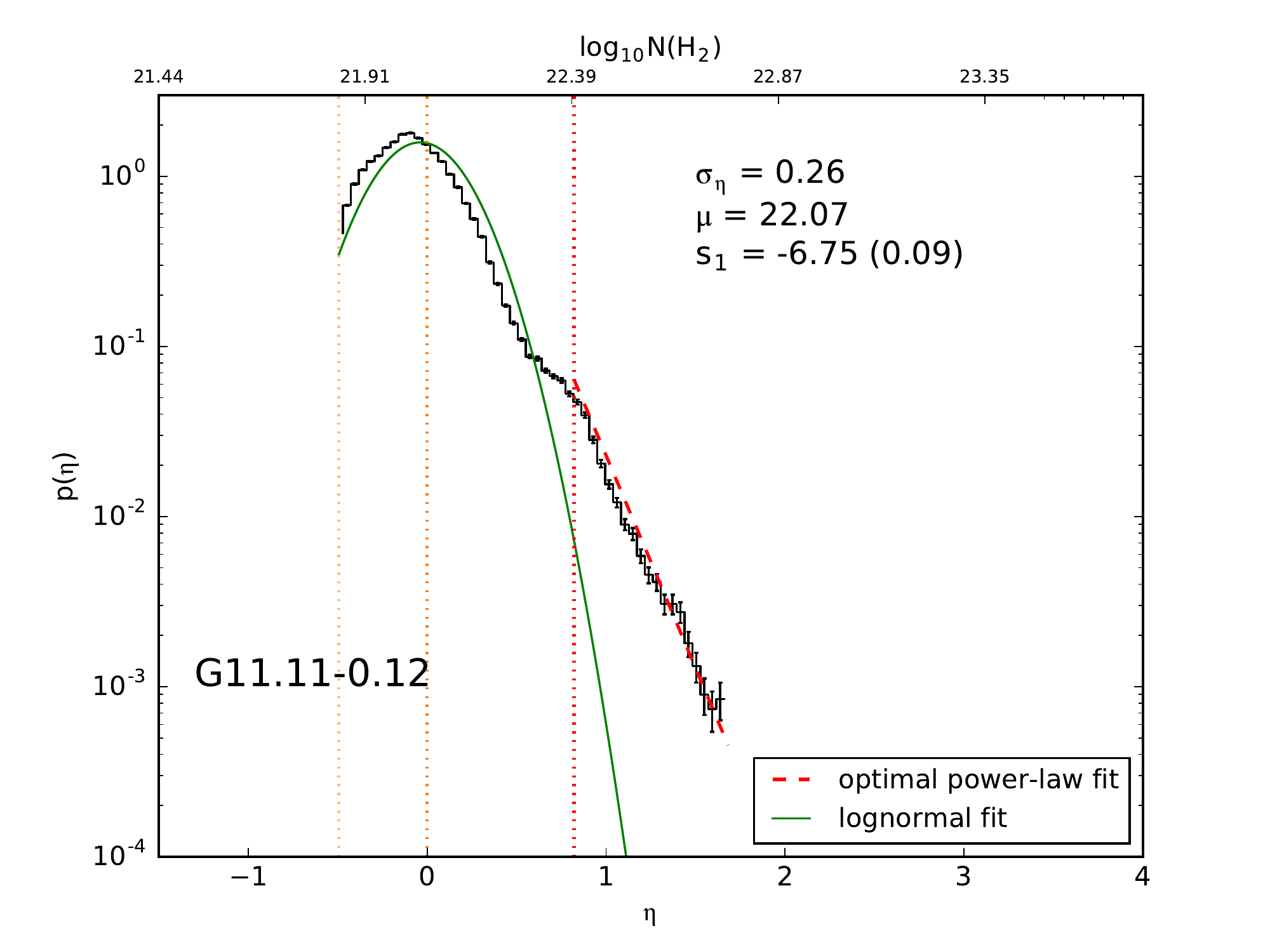}\\
\end{tabular}
\caption{Column density probability distribution functions of the observed infrared dark clouds with light orange, orange, and red vertical dotted lines indicating the measured threshold of column density, mean column density, and column density starting value of optimal power-law fit, respectively.} 
\label{fig:NPDF}
\end{figure}

\section{Results}\label{section:results}
\subsection{Molecular cloud structures and temperature distribution}\label{sub:NandTd}
Figure \ref{fig:TN_G11_hersext}, \ref{fig:TN_G14}, and \ref{fig:TN_G28} show the derived dust temperature and gas column density maps, assuming a gas to dust mass ratio of 100. For detailed calculations and procedure, see Appendix A and Lin et al. (2016).
The derived gas masses of these three IRDCs range in $\sim$3-9$\times$10$^{4}$ $\rm M_{\odot}$.
The overall gas masses and bolometric luminosities derived from these maps, are summarized in Table \ref{tab:source_info}.

The column density maps of these IRDCs present considerably different overall morphologies.
On the $\sim$30 pc scale, IRDC G11.11-0.12 appears as an integral shaped filament, which shows some wiggles on $\sim$10 pc scales.
In addition, the high angular resolution we achieve reveals that the large-scale filament may in fact consist of bundles of filaments. 
The majority of dense gas structures in G11.11-0.12 have dust temperature lower than 20 K.
We are able to identify one internally heated source embedded approximately at the center of the overall filament, where previous observations have found water masers, Class II methanol masers, and compact centimeter continuum sources (Pillai et al. 2006; Wang et al. 2014; Rosero et al. 2014).
There are several isolated heating sources, which are projected adjacent or well outside to the dense filament.
How many of those heating sources are directly associated with this molecular cloud is not yet certain.

IRDC G14.225-0.506 is resolved into a web of dense gas filaments, which seem to align in two preferred directions (see also discussion in Busquet et al. 2013).
The two most significant column density peaks in Figure \ref{fig:TN_G14} correspond to the two previously known relatively massive ($\sim$10$^{3}$ $\rm M_{\odot}$), $\sim$0.5 pc scale molecular clumps\footnote{We follow the existing nomenclature in the literature (e.g., Zhang et al. 2009; Wang et al. 2011; Liu et al. 2012a, 2012b). In this way, massive molecular clumps refer to structures with sizes of $\sim$0.5-1 pc, massive molecular cores refer to the $<$0.1 pc size structures embedded within a clump, and condensations refer to the distinct molecular substructures within a core. Fragmentation refers to the dynamical process that produces or enhances multiplicity. Molecular filaments refer to the geometrically elongated molecular structures, and molecular arms refer to segments of molecular filaments that are located within the $\lesssim1$ pc radii of molecular clumps and may not be fully embedded within molecular clumps.}.
The northern massive molecular clump, namely Hub-N, is adjacent to a compact H\textsc{ii} region, which presents a high dust temperature over a 1$'$-2$'$ angular scales.
The rest of the dense structures have $\sim$18-20 K dust temperature in general.
There are several heated sources embedded in the filamentary structures in G14.225-0.506.

The dominant dense gas structures in G28.34+0.06 can be described by a massive molecular clump in the north (P2, see Carey et al. 2000; Wang et al. 2008; Zhang et al. 2009), which is connected with a dense gas filament in the south.
The southern gas filament has several embedded clumps.

Star-formation in the P2 clump is already active.
The P2 clump is internally heated by the embedded young stars, and shows higher than 25 K dust temperature.
Dust temperature in the southern molecular clumps remains under 20 K.

\subsection{Column density probability distribution function, column density complementary cumulative distribution function, and two-point correlation function}\label{sub:statistics}
To systematically quantify the dense gas distributions in the observed IRDCs, and thereby permit comparisons with other observations and theoretical models, we perform analyses of the column density probability distribution functions (N-PDF), column density complementary cumulative distribution function (N-CCDF), and the two-point correlation functions (2PT) of gas column density.

The N-PDFs
of the three IRDCs are presented in Figure \ref{fig:NPDF}. 
For the sake of enabling quantitative comparisons with other observations, we approximate the components in the N-PDFs that have a decreasing slope at lower column densities by a lognormal distribution function, and approximate the components that have a near constant slope at higher column densities by a power-law distribution function. 
In the later discussion, we will refer to these as the ``lognormal" and ``power-law" components.
We emphasize that the functional forms we selected are merely approximations for the actual observed N-PDFs, which instead show richer features that cannot be described by a simple fit to these functional forms. 
Linking these functional forms to the underlying physics is not trivial, and one should not over-emphasize on the exact functional forms.
We elaborate further on this point in Section \ref{sub:sim} by examining comparisons of numerical hydrodynamic simulations. 

We fit an overall power-law and a power-law starting from the optimal column density cutoff based on Maximum Likelihood Estimation (MLE; Clauset et al 2009; Alstott et al. 2014) to the N-PDF of G28.34+0.06. 
The results of these two fits appear consistent. 
At the high column density end, however, the N-PDF of G28.34+0.06 shows an excess over the fitted power-law.
The N-PDFs of G14.225-0.506 
and G11.11-0.12 are better described by a lognormal component in the lower column density part, in addition to the optimal power-law fit at high column density end. Whether the lognormal component is included in the fit or not does not affect the results of the optimal power-law fit because of the MLE approach used to derive the optimal cutoff for power-law fit instead of jointly fitting the two functional forms.

The N-PDF of G14.225-0.506 presents an apparent deviation from the lognormal and power-law fits over a wide range of the normalized column density at $\eta$$\sim$1-3, and therefore it is hard to define whether there is an excess or deficit at any range of $\eta$. 
In particular, a significant ``bump'' is seen around $\eta$$\sim$2.0-2.3.
The N-PDF of G11.11-0.12 decreases the most rapidly with increasing $\eta$.
The fitted parameters are tabulated in Table \ref{tab:PDFs_fit} and are also labeled in each N-PDF panel in Figure \ref{fig:NPDF}.

\begin{table*}\footnotesize{ 
\begin{center}
\hspace{-10.5 cm}
\vspace{-0.5 cm}

\caption{Fitting results of column density probability distribution functions (N-PDFs) 
}
\label{tab:PDFs_fit}
\begin{tabular}{lccccccc}\hline\hline
Target Source &Measured Threshold & Mean Column Density&Lognormal and power-law fits\\
& log$_{10}$(N(H$_{2}))$&log$_{10}$(N(H$_{2}))$&$\sigma_{\eta}$&$\mu$&$s_{0}$&$s_{1}$\\
\hline
G28.34+0.06 &21.70&21.73&-&-&-3.98(0.01)&-3.86(0.02)\\
G14.225-0.506&21.80&22.04&0.43&21.99&-&-4.12(0.01)\\
G11.11-0.12 &21.95&22.09&0.26&22.07&-&-6.75(0.09)\\
 \hline            
\end{tabular}
\end{center}
}\end{table*}

The two-point correlation (2PT) function we used follows the same form as in Kleiner \& Dickman (1984), but instead of calculating correlation of column density fluctuations, we directly measure the correlation of column densities for each source across the observed field. The correlation strength at a separation scale (lag) of $l$ is calculated by
\begin{equation}
S_{tr}(l) = \frac{\left\langle\emph{X(\textbf{\emph{r}})X(\textbf{\emph{r+l}})}\right\rangle_{\textbf{\emph{l}}}}{\left\langle\emph{X(\textbf{\emph{r}})X(\textbf{\emph{r}})}\right\rangle_{\textbf{\emph{l}}}},
\end{equation}
where $\emph{X(\textbf{\emph{r}})}$ denotes the column density value at position \textbf{\emph{r}}, and the angle brackets are an average over all pairs of positions with a separation of $l$. The final form of the correlation function is normalized by the peak correlation strength to enable comparisons between different fields. For a more detailed description, we refer to Lin et al. (2016).

\begin{figure}
\hspace{-0.3cm}
\vspace{-0.1cm}
\includegraphics[width=9.5cm]{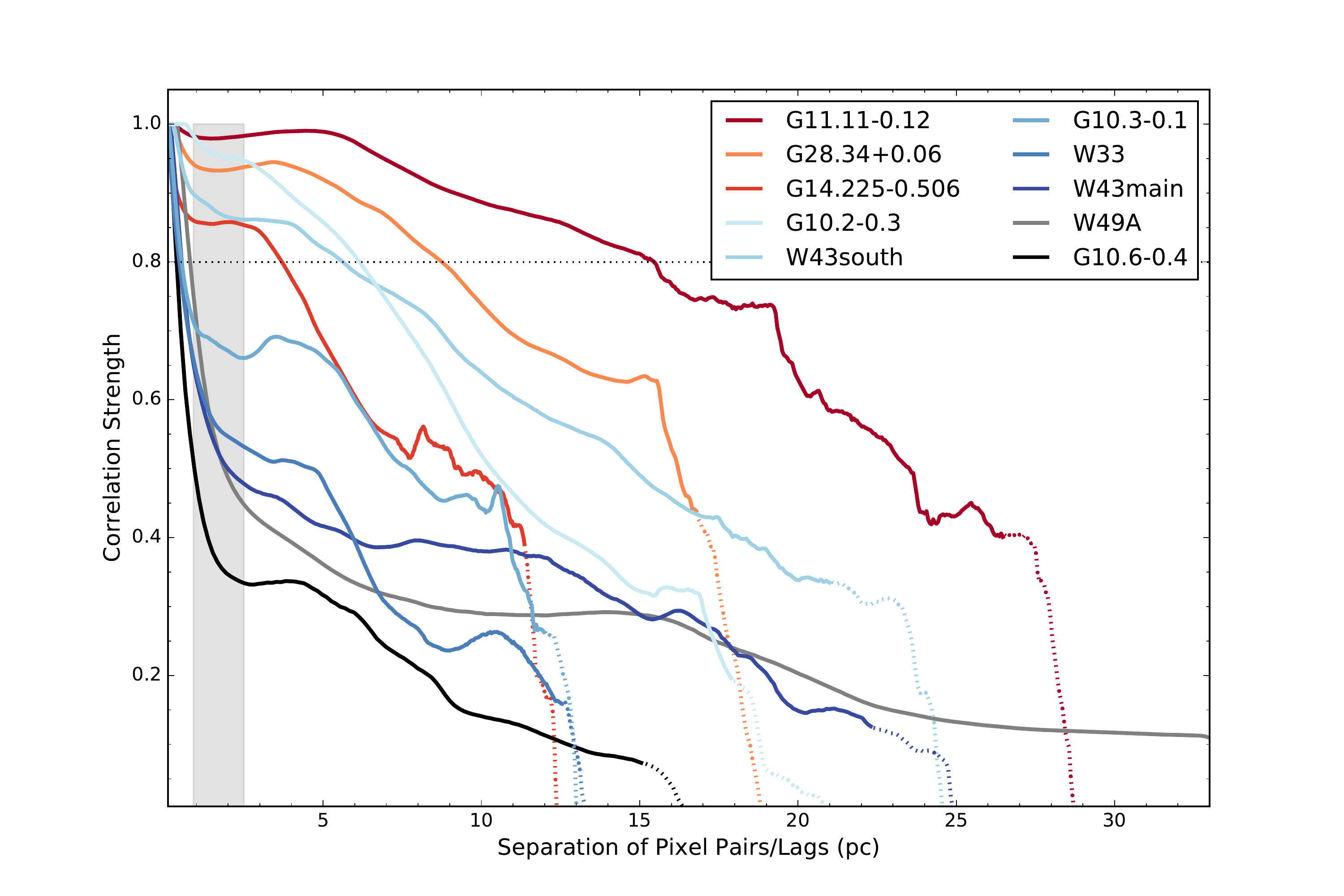}
\caption{
Two-point correlation functions of the column density distributions, for the observed sample. For separations with pixel pairs of less than $\sim90\%$ total pixel number, the correlation strengths are plotted with dashed lines. The grey filled region indicates approximately where the rapidly decreasing components end for most of OB-cluster forming regions. Black horizontal dotted line indicates correlation strength of 0.8. 
}
\label{fig:2pt}
\end{figure}

Figure \ref{fig:2pt} shows the 2PT functions of the three IRDCs overplotted with those of the active OB cluster-forming molecular clouds taken from Lin et al. (2016). 
The 2PT functions of the three IRDCs, in general, show a smooth decay of correlation strengths over all spatial scales. This is in contrast with most of the active OB-cluster forming regions which rapidly decrease at small scales. 
The dominant elongated cylindrical filaments in G11.11-0.12 and G14.225-0.506 may naturally have flattened 2PT functions since, by definition, filaments have comparable column density distributions over a wide range of separations.
For G28.34+0.06, we spatially resolved complex local over-densities scattered across the column density map. 
The multiple fluffy and scattered over-dense substructures of this molecular cloud more closely resemble the cloud morphology of W43-south and G10.2-0.3, which also have similar 2PT functions (see Lin et al. 2016).

The homogeneity of the column density fields could be estimated by the range of plateaus in the 2PT functions. 
In light of this, G11.11-0.12 would present the least mass concentration since its 2PT function remains at an almost constant correlation strength up to $\sim$6 pc. 
G14.225-0.506, which harbours two pronounced massive clumps, shows a steep decrease at lags less than 1 pc, indicating a relatively significant concentrated distribution at small scales. 
The 2PT function of G28.34+0.06 falls between those of G11.11-0.12 and G14.225-0.506 with a slight decay at small scales ($<1$ pc) and a constant correlation plateau to $\sim$4 pc.
We note that the 2PT function plateaus on the few pc scales may be related to the elongated nature of cloud structures.

\begin{figure}
\hspace{-0.3cm}
\vspace{-0.1cm}
\includegraphics[width=9.5cm]{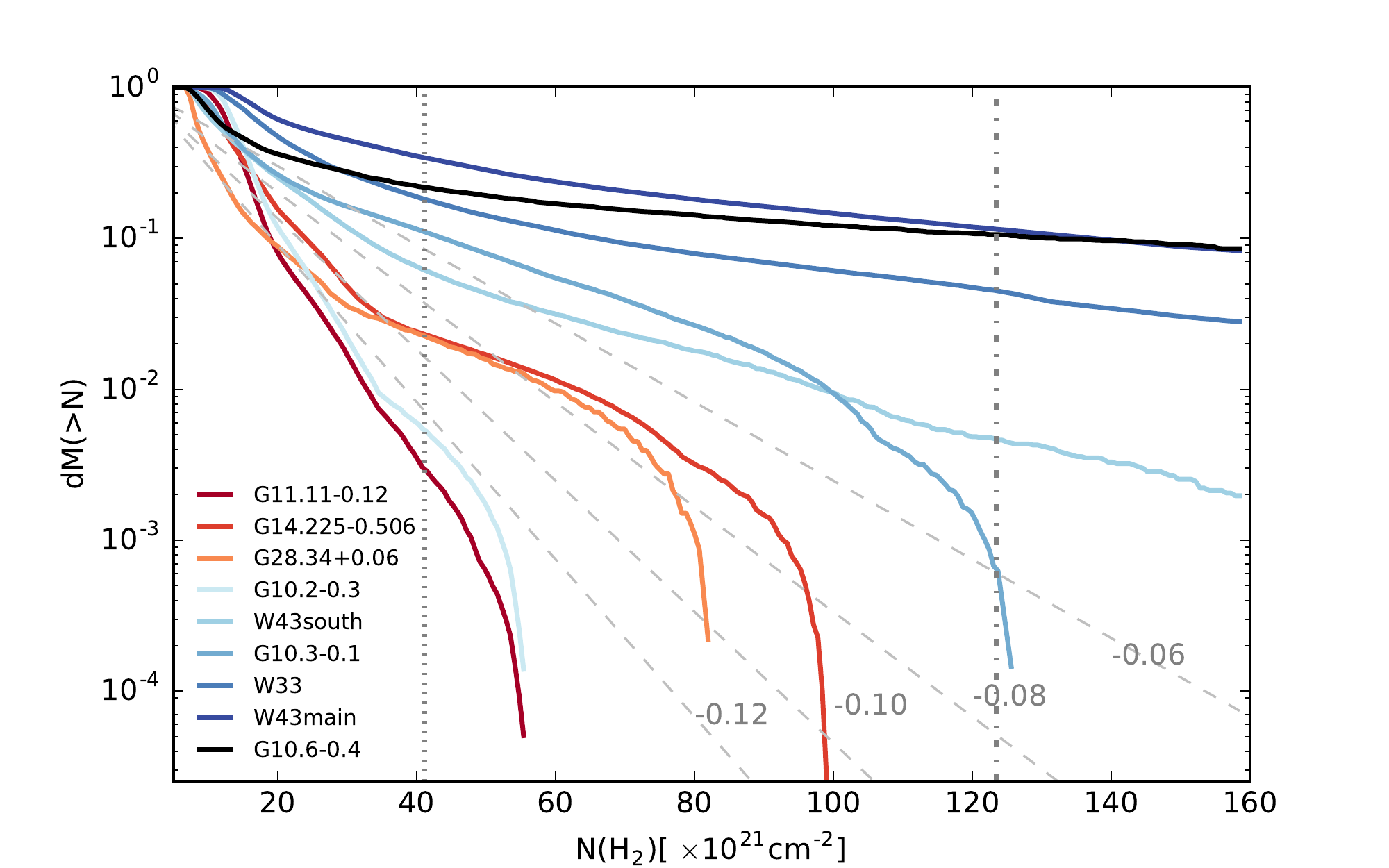}
\caption{N-CCDF of three IRDCs and OB-cluster forming regions. The lower threshold for all sources is $7\times 10^{21} cm^{-2}$. All the sources all smoothed to the resolution if they are at the distance of W43 ($\sim$5.5 kpc). The vertical dotted line indicates a core of $10^{5}$ $\rm cm^{-3}$, $\sim$0.1 pc, while the dashed dotted line indicates a core of $10^{6}$ $\rm cm^{-3}$, $\sim$0.03 pc located at same distance. Color coding for each source is the same with Figure \ref{fig:2pt}.}
\label{fig:dgmf_sm}
\end{figure}

Figure \ref{fig:dgmf_sm} presents the column density complementary cumulative distribution function (N-CCDF) of these molecular clouds (this is also defined as dense gas mass function in some previous works, see details in Kainulainen et al 2011, 2013; Ginsburg et al. 2015).
We exclude W49A, in spite of its highest fraction of dense gas, to avoid a bias from its larger distance ($\sim$11 kpc).
The dense gas fraction is closely associated with the star formation rate (SFR), raised based on observations towards both galactic and extragalactic star-forming regions (Gao \& Solomon 2004; Wu et al. 2005; Lada et al. 2010).
In Figure \ref{fig:dgmf_sm}, we plot the N-CCDFs for all the sources. The dashed grey reference lines indicate exponential decay of indexes between -0.06 to -0.12.
The N-CCDF of OB-cluster forming molecular clouds within our sample of IRDCs are generally more flattened in the high column density ends, except for G10.2-0.3 and G10.3-0.1.
These sources hold a considerable mass fraction in the high column density regime. 
Despite its relatively high total mass, G11.11-0.12 has a sharp decrease of very dense gas and is consistent with its steep N-PDF power-law tail. We additionally plot two reference vertical lines representing a 10$^{5}$ cm$^{-2}$, $\sim 0.1$pc core and a 10$^{6}$ cm$^{-2}$, $\sim 0.03$pc core lying at the same distance. 

The N-CCDF result we obtained for G11.11-0.12 is consistent with Herschel results obtained by Kainulainen et al. (2013), where the authors also use near- and mid-infrared absorption to derive a $\sim 2''$ column density map of this cloud. 
The large discrepancy of our column density map with their extinction derived one may mainly come from the resolution difference (and also the reprojection to a larger distance in our case). 
A $\sim$0.035 pc resolution extinction map enables detections of embedded dense cores which are expected to give a significant rise to N-CCDF in high column density end.
We also notice the fact that the near- and mid-infrared extinction method is limited to detecting a lower range of column density ($\rm A_{v}$$\sim$100, according to Kainulainen et al. 2013). 
Therefore, to evaluate the origin of this difference in the high column density regime one should perform a systematic error estimates of the extinction method, which is beyond the scope of our current analysis.

The sharp decrease for G28.34+0.06 and G14.225-0.506 at less than $\rm N(H_{2})$$\sim$4.0$\times 10^{22} $$\rm\ cm^{-2}$ and the near flat region up to $\rm N(H_{2})$$<$$\sim$8.0$\times$$ \rm10^{22}\ cm^{-2}$ may place these clouds at an evolutionary stage in-between
G11.11-0.12 and OB-cluster forming samples. These results imply that these two IRDCs already have a significant concentration in mass at relatively high column densities.

\section{Discussion}\label{section:discussion}

\begin{figure*}
\hspace{2.1cm}
\vspace{-0.1cm}
\includegraphics[width=14.5cm]{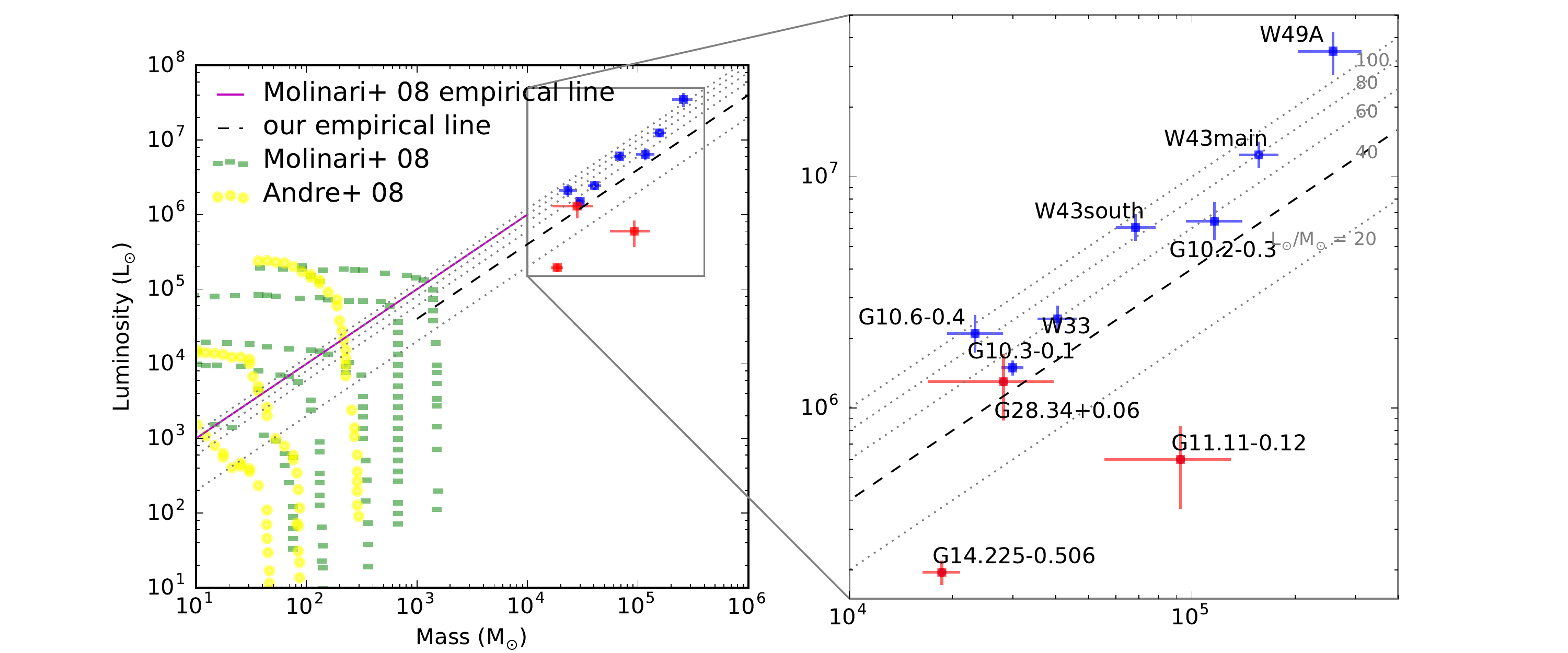}
\vspace{0.2cm}
\caption{Luminosity-mass plot of all the sources. Error bars are calculated based on uncertainties of distances. 
For the kinematic distances of G11.11-0.12 and G28.34+0.06 we adopt a distance uncertainty of 20\%. The dashed black line denotes our empirical $\rm L_{\odot}/M_{\odot} = 40$ line. Dotted reference lines are $\rm L_{\odot}/M_{\odot} = 20, 40, 60, 80, 100$, respectively. Evolutionary tracks for final (formed) different masses of massive stars are indicated in yellow dashed lines (Andre et al. 2008) and for different initial envelope masses are indicated in green dashed lines (Molinari et al. 2008). The empirical boundary line of ``Class 0" and ``Class \textsc{II}"-like $\sim$1 pc clumps derived by Molinari et al. (2008) is also indicated in the plot. Note that luminosities for OB-cluster forming regions are different from those tabulated in Lin et al. (2016) where they were calculated by summing up all pixels in the field-of-views.}
\label{fig:ttlumi_ttmass}
\vspace{0.4cm}
\end{figure*}

\subsection{Indications of physical mechanisms based on N-PDF and 2PT results}
In this section we compare the three observed IRDCs in terms of their N-PDFs results, and how these are linked to possible physical processes at work. We also compare their measured properties with observations of seven more evolved OB cluster-forming regions reported in Lin et al. (2016).
The masses and luminosities of all the sources are summarized in Figure \ref{fig:ttlumi_ttmass}.
We caution that small sample size limits our analysis to being conjecturing rather than conclusive.

The N-PDFs of G11.11-0.12 and G14.225-0.506 have a more prominent lognormal component than OB-cluster forming regions. 
Despite the possible bias induced by non-uniform sensitivity and different distances towards different sources, the N-PDF results in this paper and in Lin et al. (2016) are derived only for values greater than the measured column density threshold of each region and normalized to enable reliability in the comparisons.
The lognormal plus power-law tail N-PDFs of G14.225-0.506 and G11.11-0.12 and their highly filamentary structures resemble but are not uniquely explained by the self-gravitating cylinder model (Myers 2015). 
The relative significance of the high density part in the N-PDF of G14.225-0.506 is higher than in G11.11-0.12, with a larger $\sigma_{\eta}$ and a shallower power-law tail index $\alpha$.  The increased spatial density concentration is also seen in the sources' 2PT functions: G14.225-0.506 has a more prominent decrease of correlation strength at small spatial scales. 
Turbulent flows or shocks in the large-scale, low-column-density regime may shape the initial morphology of these clouds, producing the lognormal component in the N-PDFs (Vazquez-Semadeni 1994; Klessen 2000).
For example, Tackenberg et al. (2014) suggested that a large scale accretion flow exists along the integral filament of G11.11-0.12.
Unlike the extreme cases of W49A and G10.6-0.4, which may be undergoing global and local collapse simultaneously in the clouds, the development of power-law tails at the high column density ends of G11.11-0.12 and G14.225-0.506 may suggest that they are only gravitationally unstable over a small fraction of the observed area.

We note that the N-PDFs of these two IRDCs may not necessarily be simply understood as caused fully by the interplay between turbulence and self-gravity, considering that they both have been suggested to have dynamically non-negligible magnetic field (Santos et al. 2016; Pillai et al. 2015). 
The index of the N-PDF power-law tail and variance could be influenced by the magnetic field, as suggested by gravoturbulence simulations (Burkhart et al. 2015).
In some of these simulations, the narrower standard deviation $\sigma$ of the lognormal part of the N-PDF is found to be caused by magnetics fields acting as a cushion that sets column densities closer to their average value (Nakamura \& Li 2008; Molina et al. 2012).
The smaller variance and steep power-law tail of G11.11-0.12 seems to be compatible with these simulations.
It has been suggested that the low column density end of the lognormal component may be biased due to limited image size (e.g., Lombardi et al. 2015). 
Our measurements might be affected by the completeness issue, hence, we avoid further discussions of the lognormal component in the present research.

The relation between changes in the N-PDF power-law tails and the evolutionary stage of molecular clouds is suggested by several recent observational and numerical simulation works. 
For example, Stutz \& Kainulainen (2015) investigate the variations in the N-PDF slopes with young star formation content in the Orion molecular cloud, and they find that the N-PDF slope is steeper in regions where there is a smaller fraction of Class 0 protostars. 
As the power-law tail flattens, the fraction of young protostars increases.
The shallower power-law tail related to more active star formation is also suggested by Lombardi et al. (2015) based on investigations towards several molecular clouds.
The possible relation between the power-law tail slopes and collapsing state of molecular clouds is also raised by Kritsuk et al. (2011), Ballesteros-Paredes et al. (2011), Burkhart et al. (2015) etc. in their simulations. 
Regarding cloud properties reflected by N-CCDF measures, Kainulainen et al. (2013) find that the slope of N-CCDF relies sensitively on the turbulence driving mechanism and SFE, and magnetized simulations of clouds have steeper slopes than non-magnetized ones.  
The 2PT functions of the selected IRDCs show slowly decaying correlation strengths at small spatial scales, indicating less matter concentration in local areas as compared with most of the more luminous OB-cluster forming regions. On the other hand, the IRDCs  exhibit a larger decrease of correlation strengths at larger scales.
The different (mixed) driving modes and magnetized nature of the turbulent motions in these IRDCs, compared to those in luminous OB cluster-forming regions, may be the origin of the above mentioned differences in our statistical measurements (Federrath et al. 2009).

The N-PDFs of G11.11-0.12 and G28.34+0.06 were also derived by Schneider et al. (2015) based on Herschel data only.
Despite their different angular resolution, the overall shape of N-PDFs are similar, while we are able to resolve more localized dense structures.
We note that the main caveat of utilising N-PDF features to quantify cloud properties is that it is an indirect indicator of cloud physical state, unlike, i.e., volume density profiles. It is possible to convert column density maps to densities using simple assumptions of cloud geometry (Stutz \& Gould 2016), but this is not readily applicable to our sources considering their complex morphological structures based on previous spectroscopic works which show they have multiple velocity components.

In Figure \ref{fig:ttlumi_ttmass} we plot the total luminosity vs. mass of these sources with reference lines corresponding to constant luminosity-to-mass ratios.
The luminosity-to-mass ratio provides a measure of evolutionary state.
The ratio places G28.34+0.06 in a similar stage with {some of the OB-cluster forming regions, whereas G14.225-0.506 and G11.11-0.12 are more quiescent regions.
The slope of the N-PDF and the luminosity-to-mass ratio of the compared sources are presented in Figure \ref{fig:ttlumittmass_pdfslope}. 
The data points in the plot suggest a correlation between these two quantities. 

\subsection{A comparison with simulations of luminous OB cluster-formation}\label{sub:sim}

We here compare our observations with recent numerical simulations by Dale et al. (2015), who followed the formation of OB clusters from molecular clouds, and examined the impact of the clusters' feedback on the clouds. 
The simulations were performed in the Smoothed Particle Hydrodynamics (SPH) formalism. 
Smooth spherical clouds of a range of masses were initialised with a turbulent velocity field and allowed to evolve and form stars. Once a few massive stars had formed in each cloud, the effects of their winds and ionising radiation were modelled. 
The evolution of the clouds followed as close as was practicable to the epoch when the massive stars were expected to explode as supernovae, allowing the influence of the winds and radiation to be isolated.

We choose the $10^{4}$\,M$_{\odot}$ Run I calculation from Dale et al, since it is among those with the best linear resolution. 
We compute simple column-density images along the simulation $z$--axis of 10$\times$10\,pc subfields of the cloud, comparable to the physical sizes of our observed IRDCs (see also Appendix B). 
Column-density images are produced at four epochs \textit{prior to} the formation of stars and the initiation of feedback (top row of Figure \ref{fig:simn}), at the point when feedback begins to act (first panel in the bottom row of Figure \ref{fig:simn}), and for three later epochs (remaining panels in bottom row of Figure \ref{fig:simn}).

The four images of top row in Fig \ref{fig:simn} 
show the development of self-gravitating filamentary structures from the cloud's turbulent velocity field before the onset of star-formation. 
The gaseous filaments are initially not gravitationally unstable and, at early stages, their main role is to feed gas to a forming, parsec-scale massive molecular clump (the bright compact object located in the lower left corner of the two rightmost images in the top row). 
This quickly accumulates a large amount of mass and becomes the earliest site of star formation. 
In Figure \ref{fig:simnpdf}, we compute N-PDFs for these column-density images in the same way as we did for the observational data.
The absence of many massive molecular clumps in the simulated cloud is due to the initial dominance of global collapse.

At the earliest epoch when the filamentary structure is least well-defined, the PDF has a form close to lognormal in appearance. 
We caution that the drop-off at low $\eta$ is likely an incompleteness effect caused by the limited field of view and the lowest density contours not being closed.
As the simulation progresses and the filaments become self-gravitating in preparation for forming stars, the N-PDFs develop pronounced power-law tails which extend to higher and higher values of $\eta$, as do the observed plots (Figure 10).
The slopes of the power-laws are initially very steep, but flatten with time, approaching the slopes in the observed N-PDFs.

In even more striking resemblance to the observations of G14.225-0.506 (Figure \ref{fig:NPDF}), the simulated N-PDFs also shows a similar bump in the simulated data at $\eta\approx2-2.5$ (b, c, d in Figure \ref{fig:simn}). This is dominated by the protocluster core mentioned earlier, indicating that material piles up in this structure for some time, slowing its progression to still higher densities and resulting in a bump in the N-PDF. There is also a minor contribution to the bump from the ambient dense gas filaments which are immediately feeding the massive molecular clump.
The massive molecular clump rapidly contracts due to self-gravity, which produces internal structures with high density and high column density. It then no longer appears as a bump in the N-PDF, whose slope extends further to higher values (e, f in Figure \ref{fig:simn}). The N-PDF evolves to a broken power-law shape at larger than zero $\eta$, showing an excess of high column density pixels.
Such an N-PDF is similar to the observed in G28.34+0.06, and those of some active OB cluster-forming regions (e.g., G10.6-0.4, W49A) reported in Lin et al. (2016).

In our simulations, the lognormal like N-PDF is thus \textit{not} smoothly lifted to become a power-law like N-PDF, but instead changes via a rather dynamical process which produces non-smooth and non-steady features.
The bottom panels (e,f,g,h) of Figure \ref{fig:simn} shows the reaction of the N-PDF to feedback from the OB cluster which eventually forms in the dense core. The massive molecular clump forms a condensed cluster of stars; some stars are ejected due to many-body gravitational interactions.
Subsequently, the massive clump is destroyed by feedback and by consumption due to stellar accretion.
Feedback of the stars formed in the massive clump then quickly disperse the low density gas and destroys the filamentary structure from which the clump originally formed.
As a result, a cavity is gradually created at the center of the system. The action of feedback is initially rather modest and is essentially to drive the high-density tail to higher and higher values of $\eta$. The N-PDF at this stage shows a deficit of high column density gas when compared with a single power-law fit (g in  Figure. \ref{fig:simn}), which is qualitatively similar to the observed N-PDF of the evolved OB cluster-forming region G10.2-0.3 (Lin et al. 2016).
However, G10.2-0.3 presents several H\textsc{ii} bubbles, which may shape the N-PDF differently. 
Eventually, the power-law slope of high column density gas becomes steeper again (h in Figure. \ref{fig:simn}), as feedback becomes intense enough to significantly shape the high column density end.
The smoothed N-PDFs shown in Fig. \ref{fig:simnpdf} are simulated images smoothed to the resolution of our achieved resolution 10$''$ at a median distance of our sources of 3 kpc. The smoothed N-PDFs and their fitted optimal power-law tail slopes are close to the original ones except they exhibit clear truncations, which means that smoothing and differing distances do not affect our conclusions significantly but that high resolution column density maps are essential to robustly recover the suppressed power-law tails at high column density end.

As discussed in detail in Dale et al, feedback in this cloud is able to generate dense gas by compressing the outer regions of the cloud into shells, but much of this gas fails to become gravitationally unstable because it is located far from the cloud's main potential well, and is stablised by geometrical stretching and turbulence.
We also examined one of the larger clouds modelled by Dale et al (2015, the 3$\times$10$^{5}$ $\rm M_{\odot}$ UUCL cloud). We  obtained qualitatively similar results to those obtained for Run I, although the coarser mass and length resolution of Run UUCL does not allow the high-density end of the PDFs to be resolved as is possible in Run I.

\subsection{Characterizing cloud evolutionary stage based on cloud structures and temperature/luminosity profiles}\label{sub:evolution}
In the field of high-mass star-formation, the evolutionary stages of massive (e.g., $\rm M_{gas}$$>$10$^{4}$ $\rm M_{\odot}$) molecular clouds are commonly separated into the {\it infrared dark stage}, the {\it active star-forming stage}, and the {\it dispersed stage}. 
Infrared dark clouds are considered to be the youngest evolutionary stage,  representing the initial conditions of high-mass star-formation.
However, unlike low-mass stars, high-mass stars form in much shorter (Kelvin-Helmholtz cooling) timescales than the characteristic (free-fall) timescales for molecular cloud evolution.
Therefore, characterizing molecular cloud evolution based on the illumination from embedded high-mass stars, may be coarse and biased.
In terms of terminology, whether a molecular cloud is infrared dark or not, may also depend on the sensitivity and the field of view of the observations, and the time variability of the observed stellar emission, which can be ambiguous (e.g. Feng et al. 2016). 
\begin{figure}
\hspace{-0.3cm}
\vspace{-0.1cm}

\includegraphics[width=9.5cm]{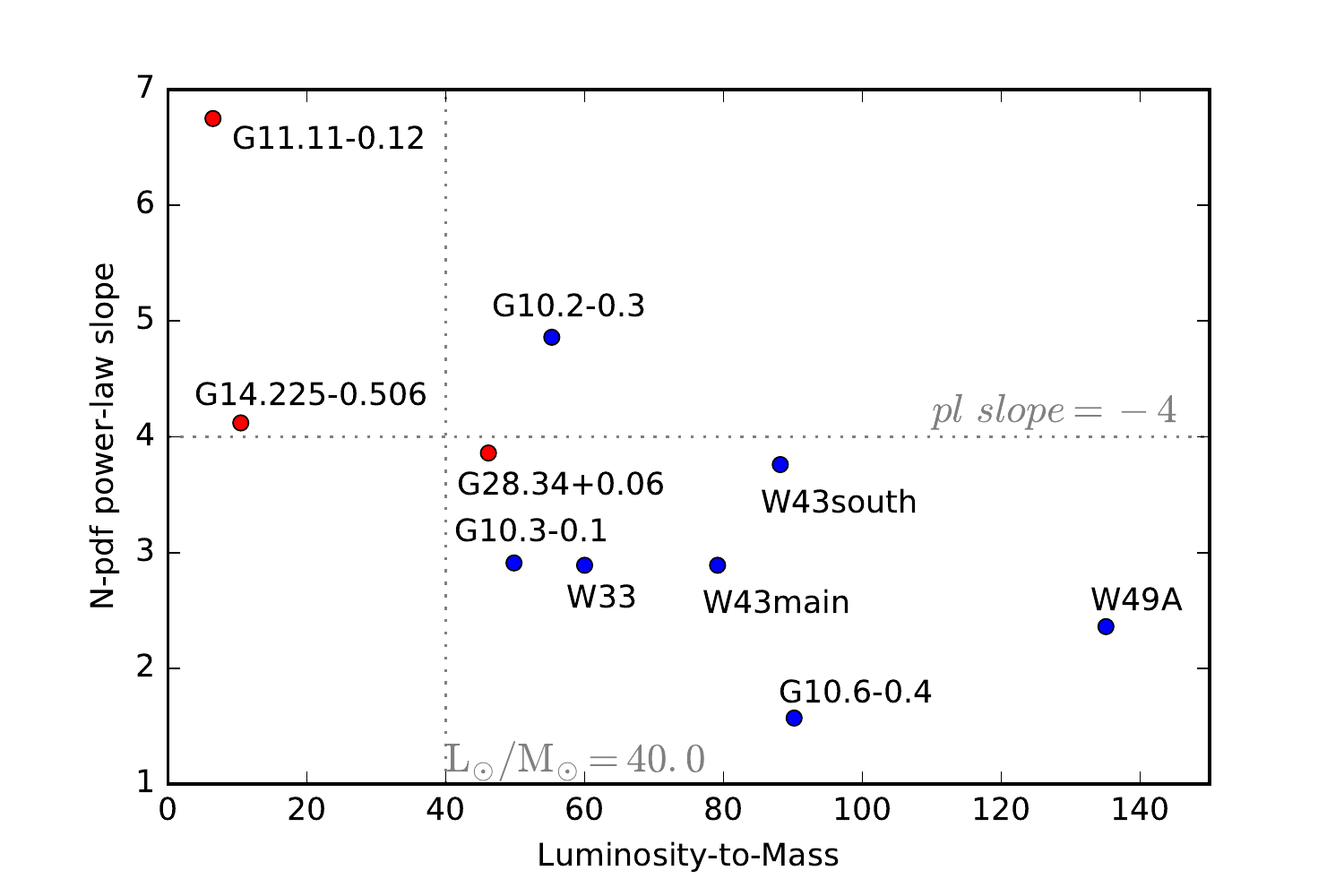}
\caption{Luminosity-to-mass ratio vs. slope of N-PDF power-law slope at high column density for all the sources. Luminosities and masses are in solar units.}\label{fig:ttlumittmass_pdfslope}
\end{figure}

\begin{figure*}
\begin{tabular}{p{4cm}p{4cm}p{4cm}p{4cm}}
\includegraphics[width=4.5cm]{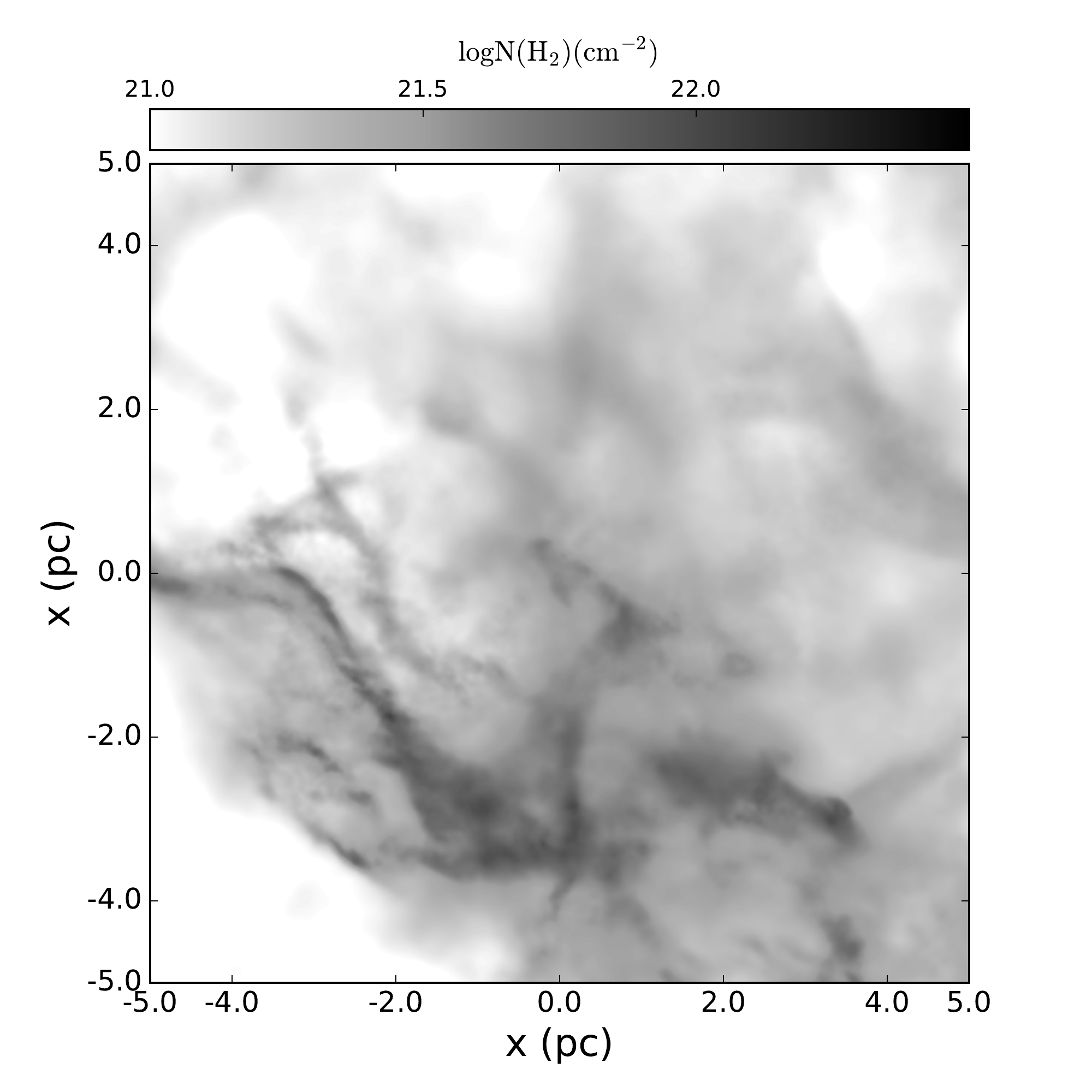} &  \includegraphics[width=4.5cm]{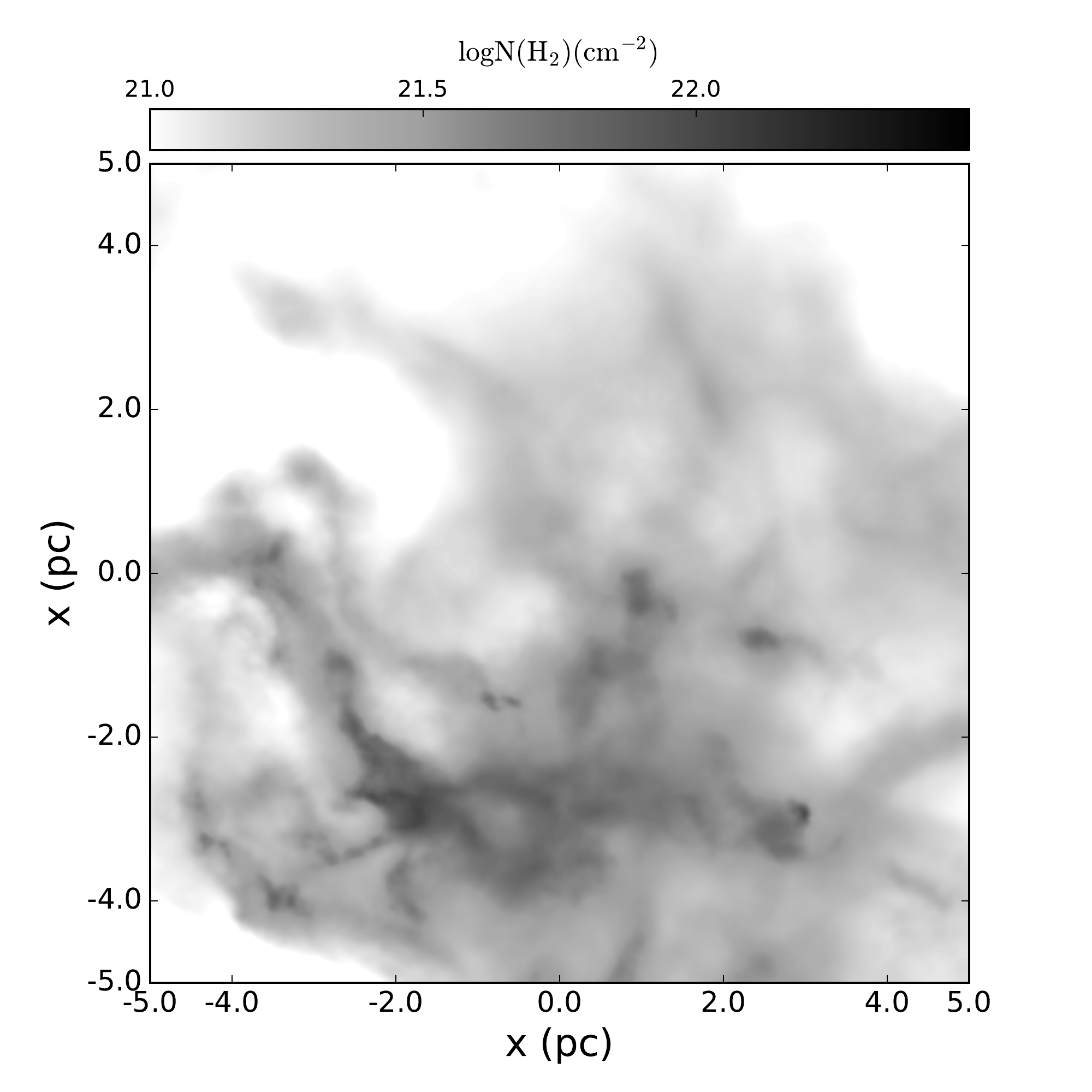} &
\includegraphics[width=4.5cm]{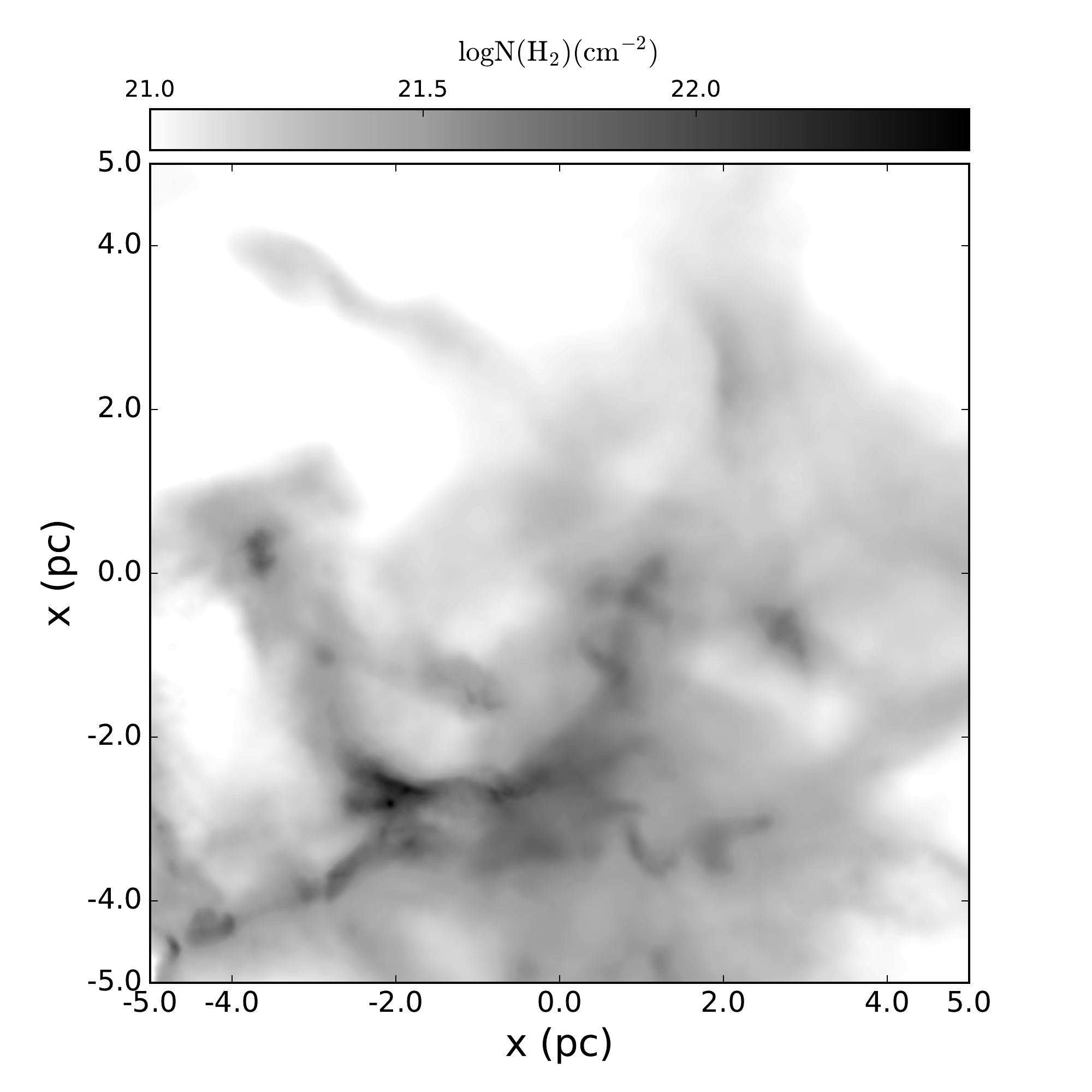} & 
\includegraphics[width=4.5cm]{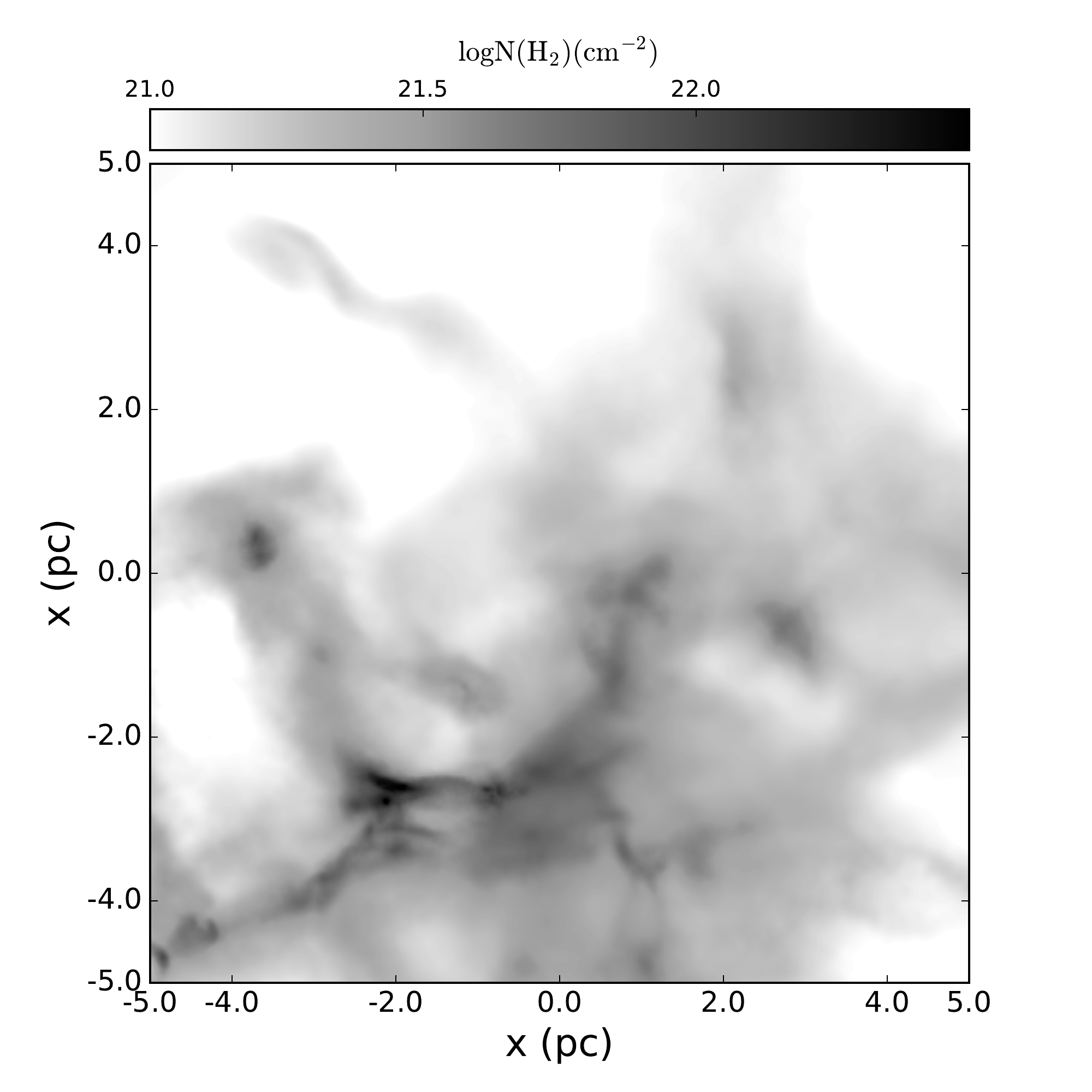} \\
\includegraphics[width=4.5cm]{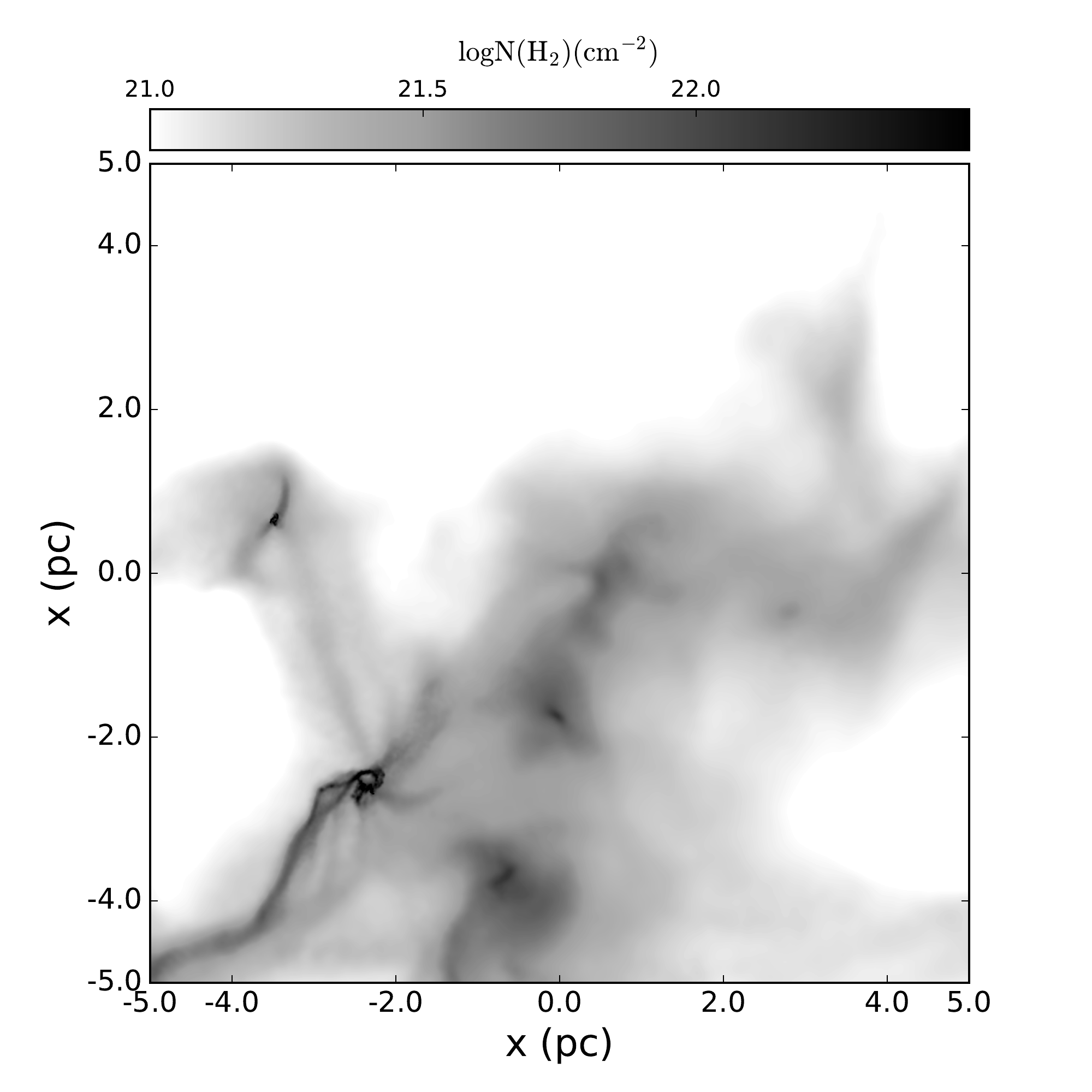} &  \includegraphics[width=4.5cm]{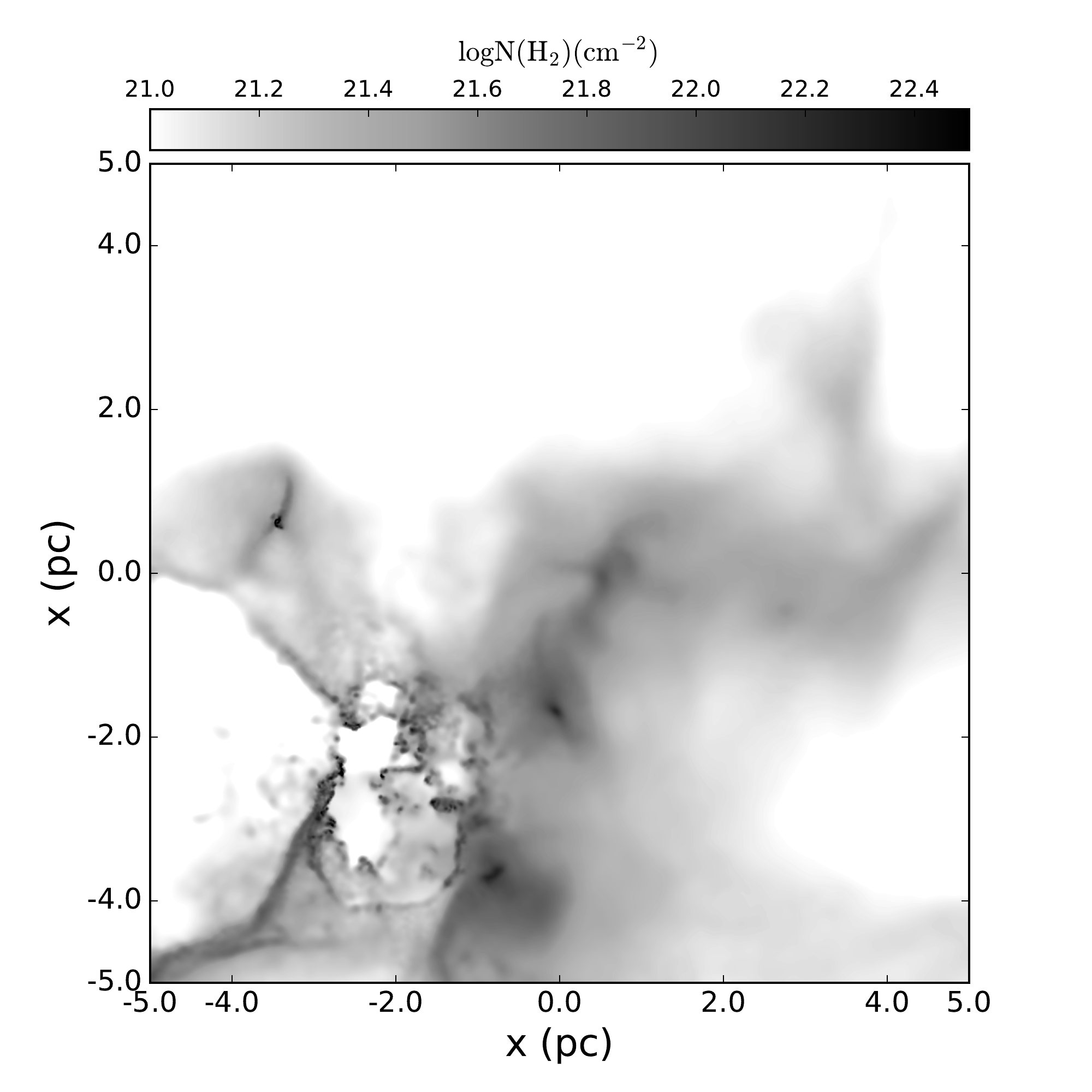} &
\includegraphics[width=4.5cm]{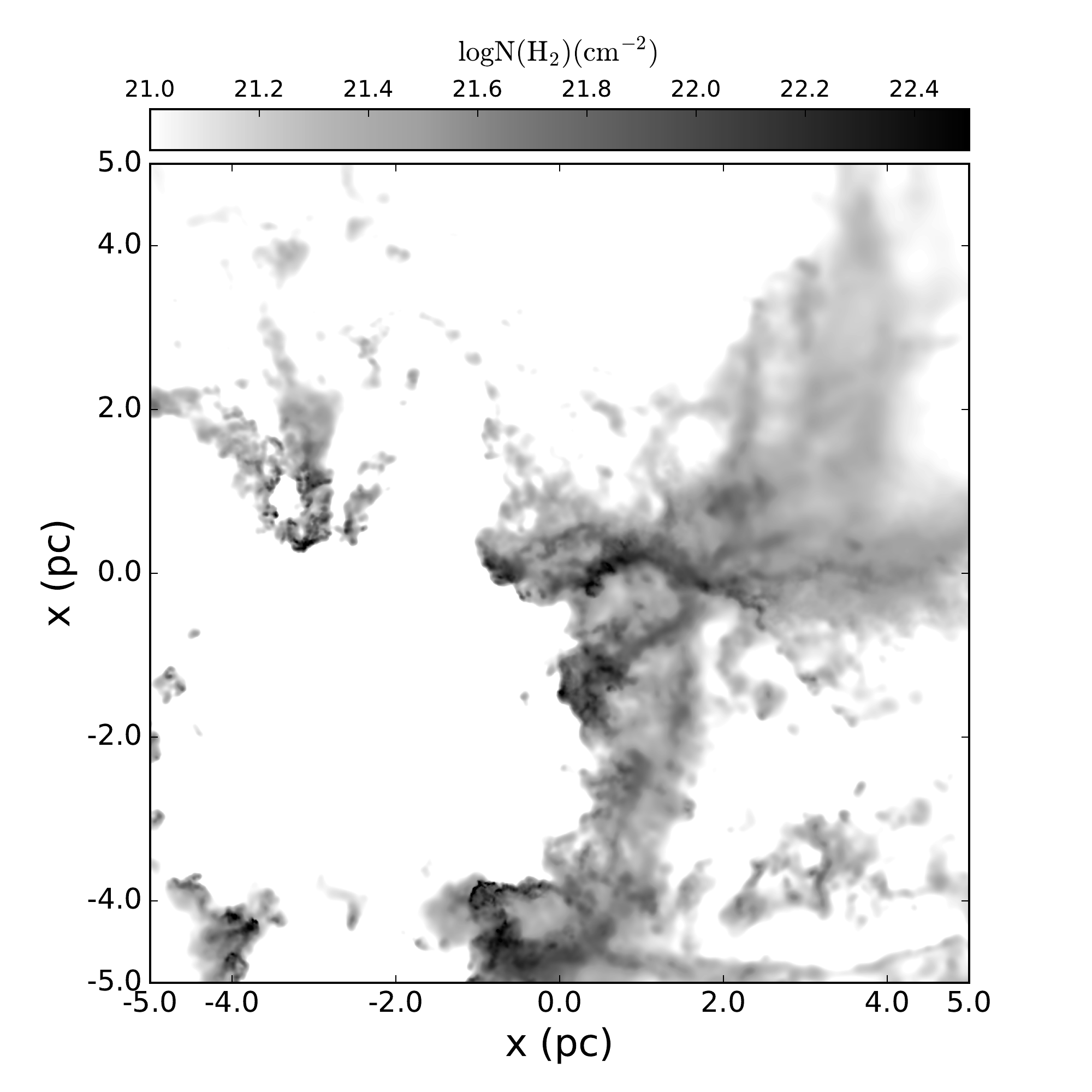} & 
\includegraphics[width=4.5cm]{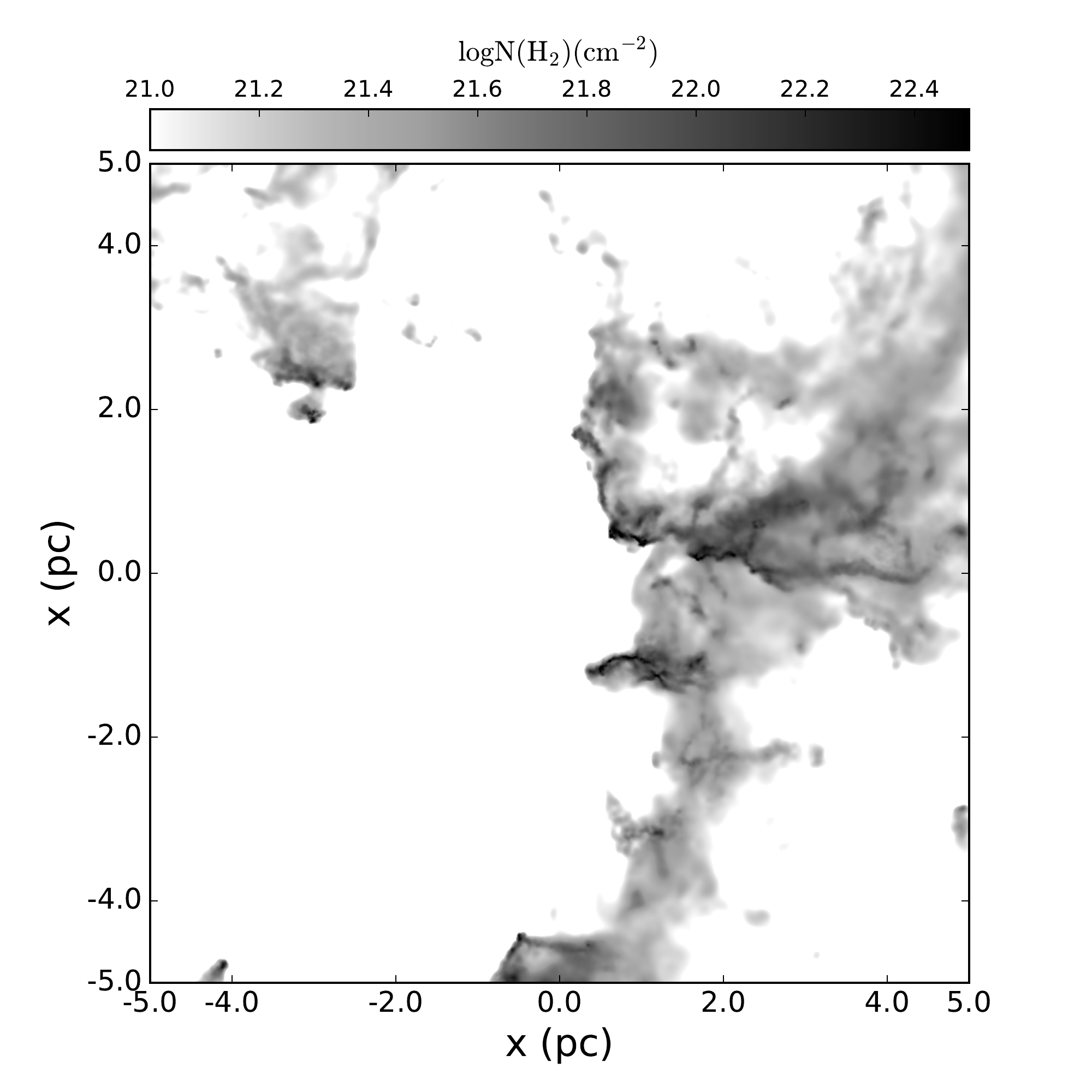}  \\
\end{tabular}
\caption{Simulated column density evolution of a 10$^{4}$ $\rm M_{\odot}$ molecular cloud.
Panels in the top row (from left to right), which are before the onset of star-formation, show the time epoch of 2.4 Myr, 3.2 Myr, 3.7 Myr, and 3.9 Myr, respectively. 
Panels in the bottom (from left to right) show the time epoch of 5.4 Myr, 5.5 Myr, 6.0 Myr, and 6.5 Myr, respectively. 
Color scale of the first 5 epochs is the same with our IRDC column density map.
More descriptions are in Section \ref{sub:sim}.
The N-PDFs are summarized in Figure \ref{fig:simnpdf}.}
\label{fig:simn}
\end{figure*}

Establishing  {\em evolutionary tracks} of massive molecular clouds based on their structures may be more essential, and crucial for facilitating studies of the co-evolution of the (proto)stellar and core mass functions with the molecular cloud (e.g., Zhang et al. 2015).
Based on a limited sample (the present work and Lin et al. 2016), and also based on the results of hydrodynamic simulations (Section \ref{sub:sim}), we hypothesize four common characteristic evolutionary stages of massive molecular clouds.
From the earlier to the later evolutionary stages they are (1) {\it cloud integration stage}, (2) {\it stellar assembly stage}, (3) {\it cloud pre-dispersal stage}, and (4) {\it dispersed cloud stage}.

In the {\it cloud integration} stage, high column-density (e.g., $N(H_{2})$ $\ge$10$^{22}$ cm$^{-2}$) gas structures are forming on  $>$10 pc scales. 
At this stage, the N-CCDF remains steep. 
The gas/dust temperature may be $<$20 K and the gas temperature is, in general, anti-correlated with the gas/dust column density, despite a few hot spots probably associated with deeply embedded intermediate-/high-mass (proto)stars.
Cloud structures at this stage are characterized by a steep and decreasing slope of the power-law tail in the N-PDFs, and the diffuse nature of the 2PTs do not show a significantly distinguishable component in the short lags.  
In $\sim$10 pc scales, the low column density part of the N-PDFs may be described by lognormal distributions, until the self-gravitational contraction outweigh other physical mechanisms present from the initial conditions (e.g., turbulence, magnetic field, etc). 
In addition, the N-CCDF and the luminosity-to-mass ratios are both low.

In the {\it stellar assembly stage} stage, the slopes of the power-law tail in the N-PDFs become shallower due to the overall self-gravitational contraction of the molecular cloud.
In addition, the highest column density end of the N-PDF may start to present an excess of single power-law tail
due to the presence of very massive molecular clumps/cores, which are forming the highest mass stars of the final clusters.
The 2PT functions at this stage start to present a distinguishable component at lags$\lesssim$1 pc, which is related to the presence of the above mentioned very massive molecular clumps/cores.
The luminosity-to-mass ratio at this stage increases very rapidly due to the quickly forming high mass stars.
From the two dimensional histogram of the gas/dust temperature and column density, we can also identify positive correlations from the core/clump material of the immediate surroundings of the newly formed high-mass stars.

In the {\it cloud pre-dispersal} stage, the remaining dense gas associated with massive molecular clumps/cores is either accreted onto the young massive star(s) (Ginsburg et al. 2016), or dispersed by the stellar feedback (Dale et al. 2015).
The distinguishable component at the $\lesssim$1 pc lags of the 2PT functions become less prominent or it fully disappears.
The slope of the power-law tail in the N-PDFs remains shallower than that in the {\it cloud integration} stage.
The luminosity-to-mass ratio and the dense gas fraction at this stage will remain high as well. 
It is until the {\it dispersed cloud stage} that  radiative feedback and the expansion of the ionized gas become capable of reverting  the contracting motions on all spatial scales, significantly re-shaping molecular gas structures. 
The N-PDF becomes steeper back again and may show an increasing slope towards high column density end.  
This evolutionary stage can be easily distinguished from the {\it cloud integration} stage and other evolutionary stages with similar N-PDF slopes because the gas/dust temperatures are much higher, 
and the presence of extended mid infrared and radio continuum emission, which trace the extended and expanding H\textsc{ii} regions.
Whether the N-PDFs can be reverted by the feedback or not, may also depend on the initial cloud morphology (e.g., the cross section against feedback).
In the case that the stellar feedback is only strong enough to disperse low density gas, the piled up high density gas may re-collapse in some free-fall timescales, re-cycling again.

\begin{figure*}
\hspace{-0.4cm}
\begin{tabular}{p{4.2cm}p{4.2cm}p{4.2cm}p{4.2cm}}
\includegraphics[width=5cm]{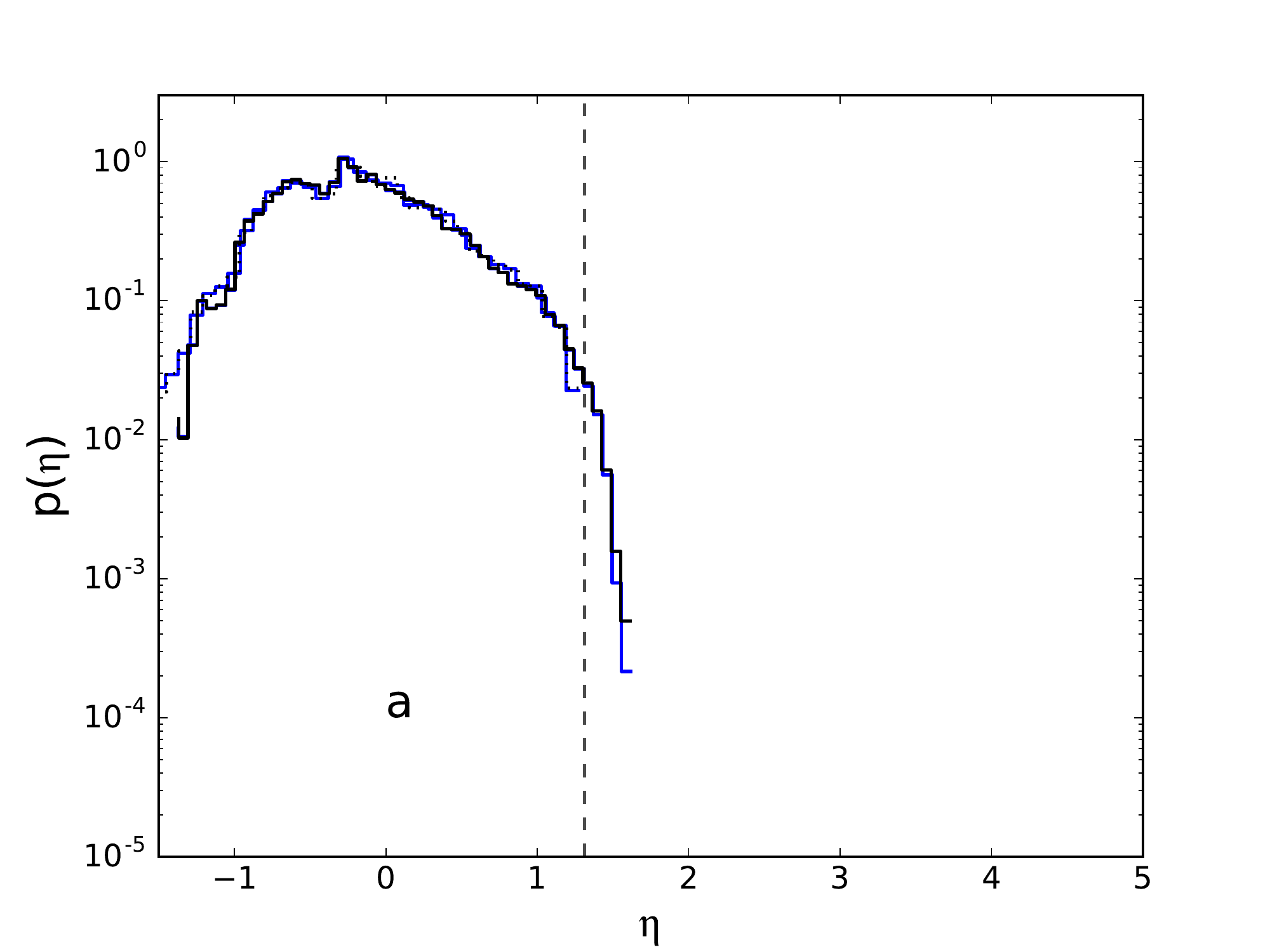} &  \includegraphics[width=5cm]{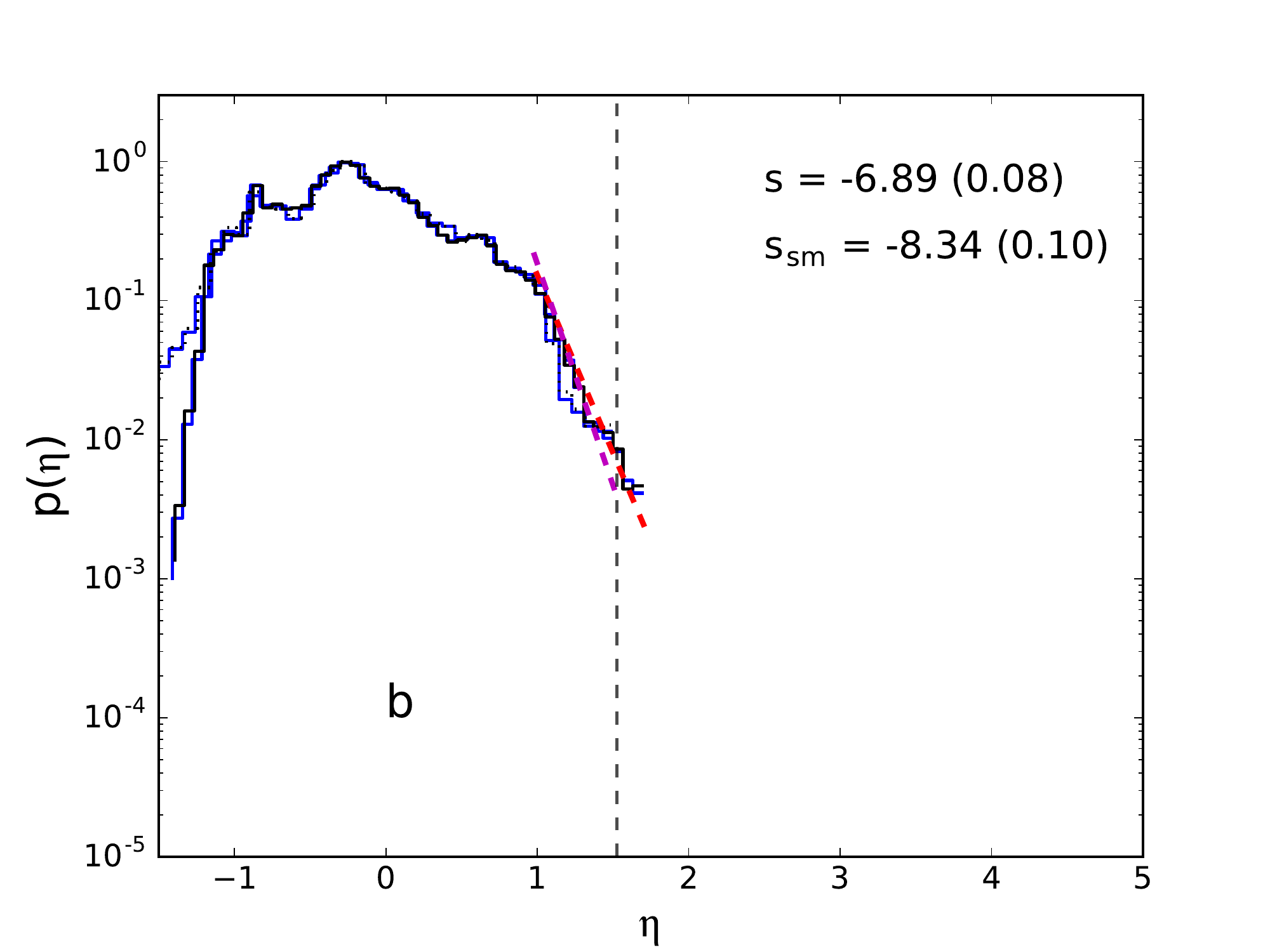} &
\includegraphics[width=5cm]{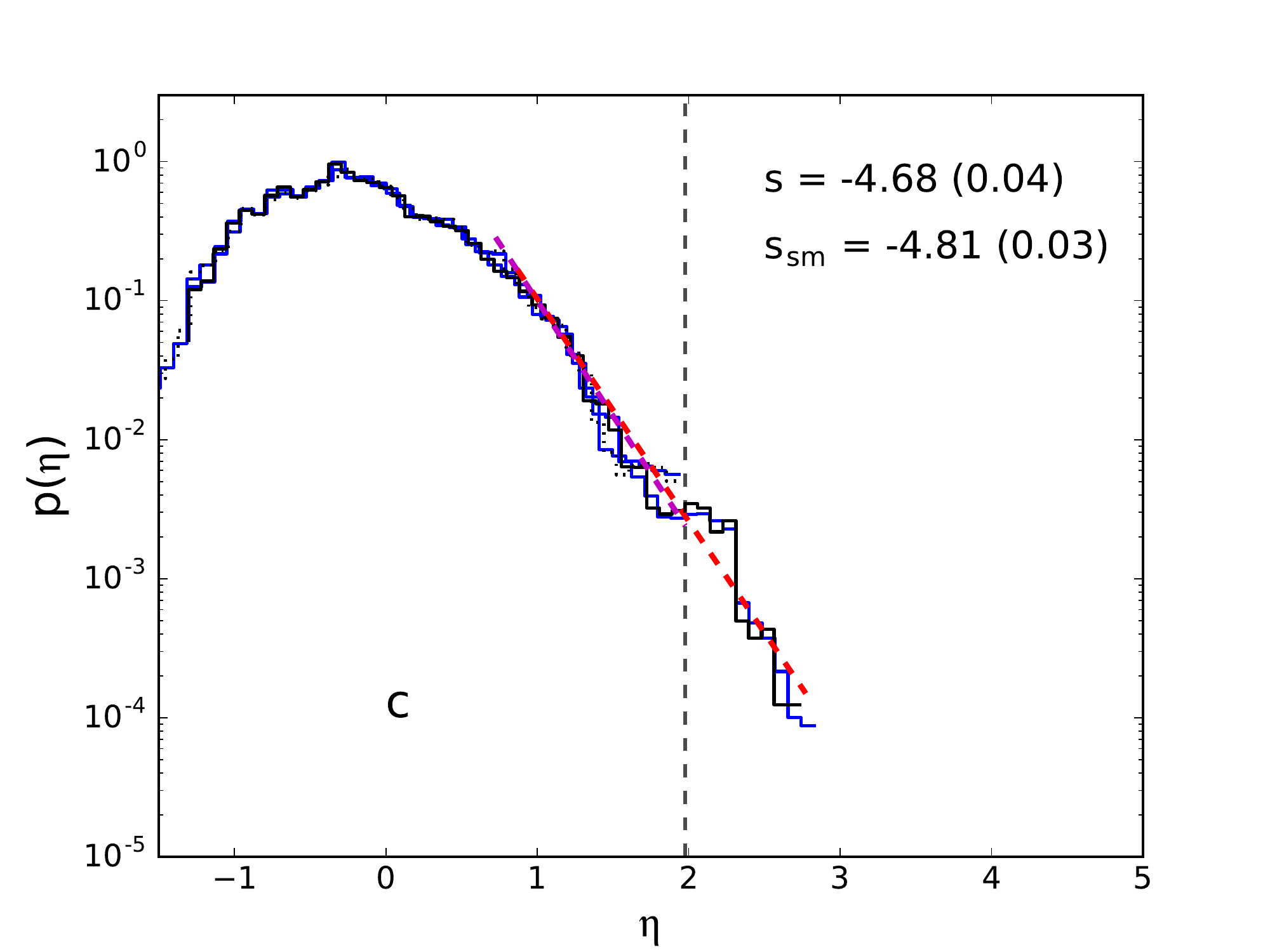} & 
\includegraphics[width=5cm]{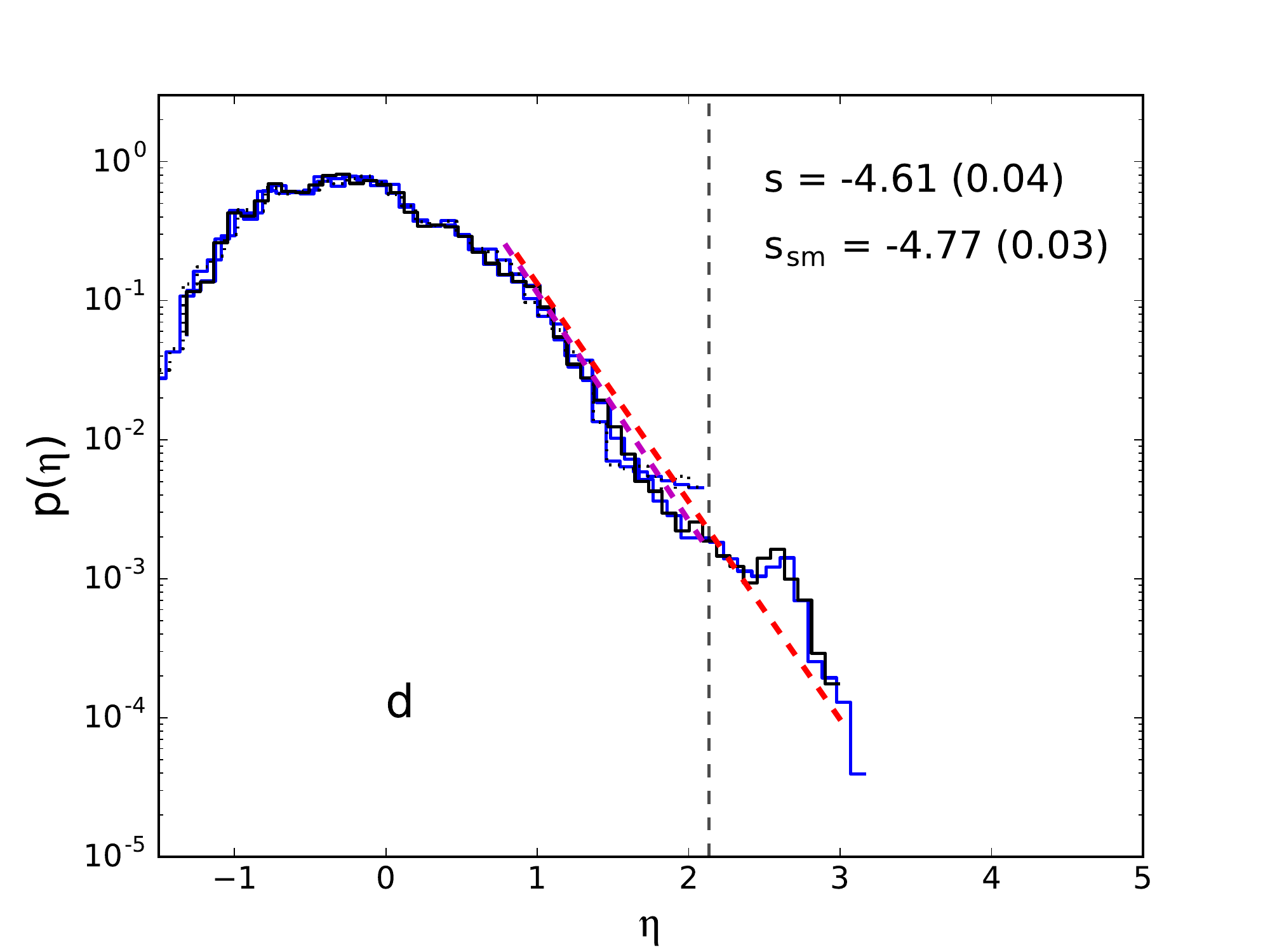}  \\
\includegraphics[width=5cm]{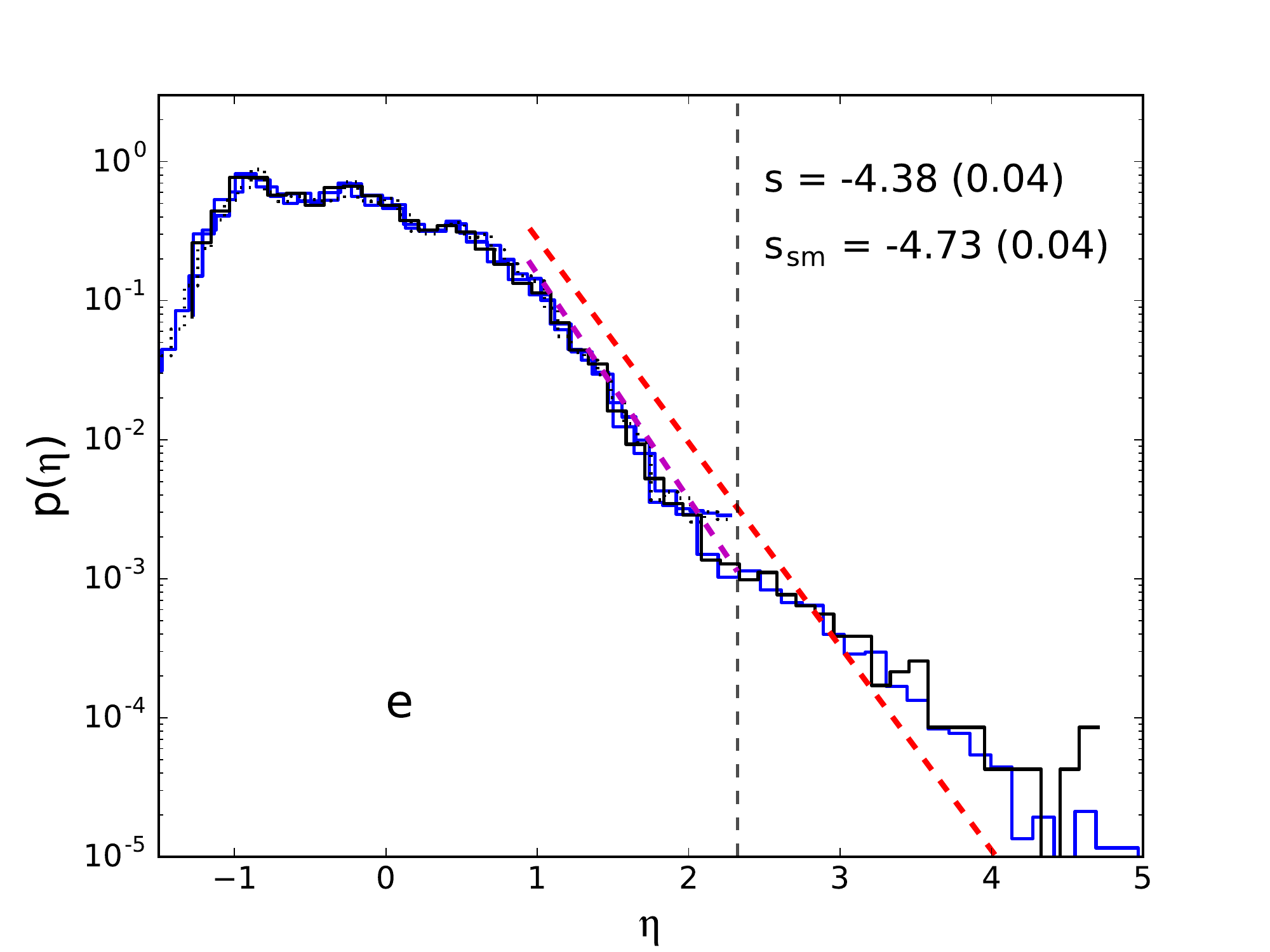} &  
\includegraphics[width=5cm]{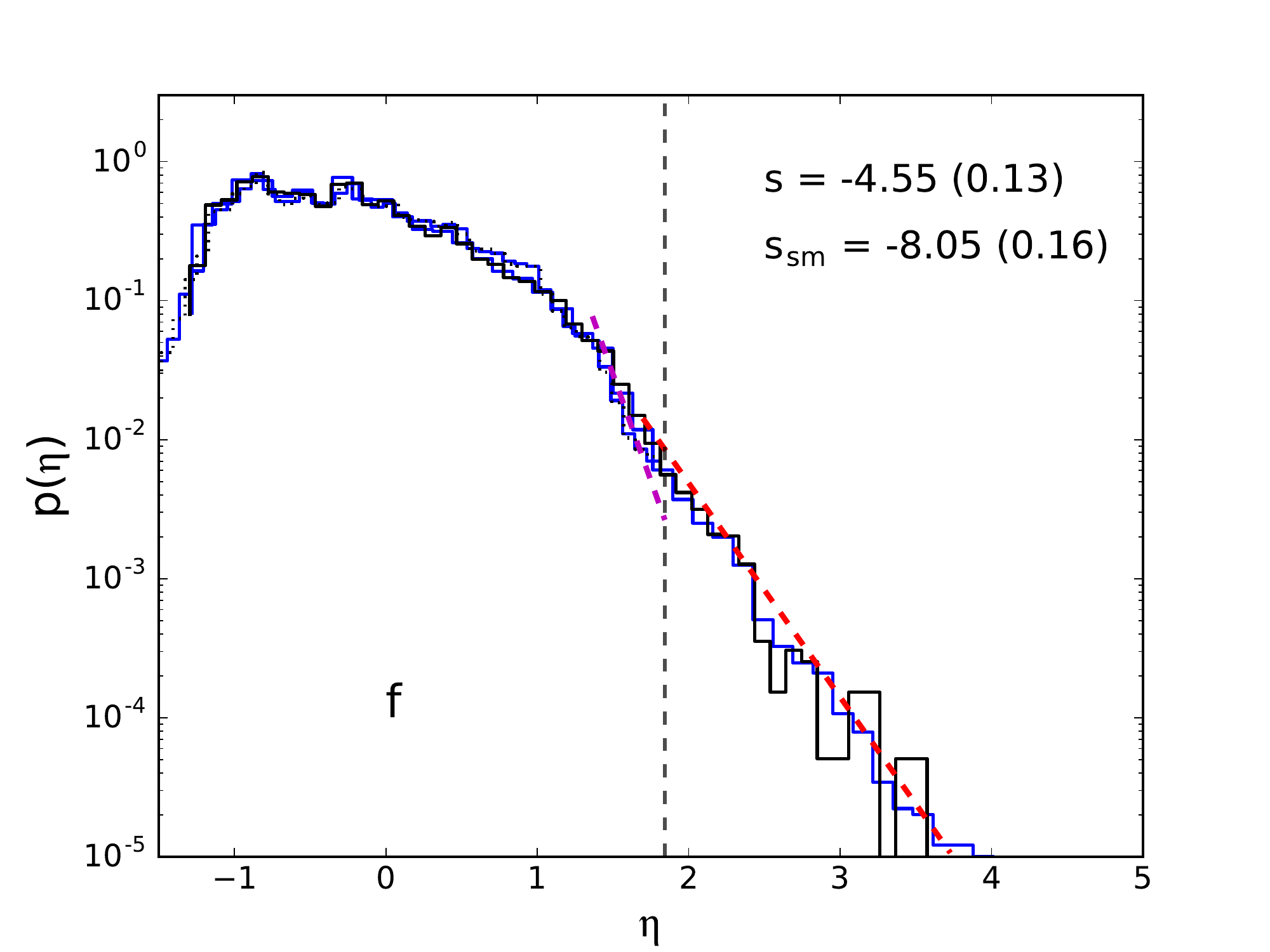} &
\includegraphics[width=5cm]{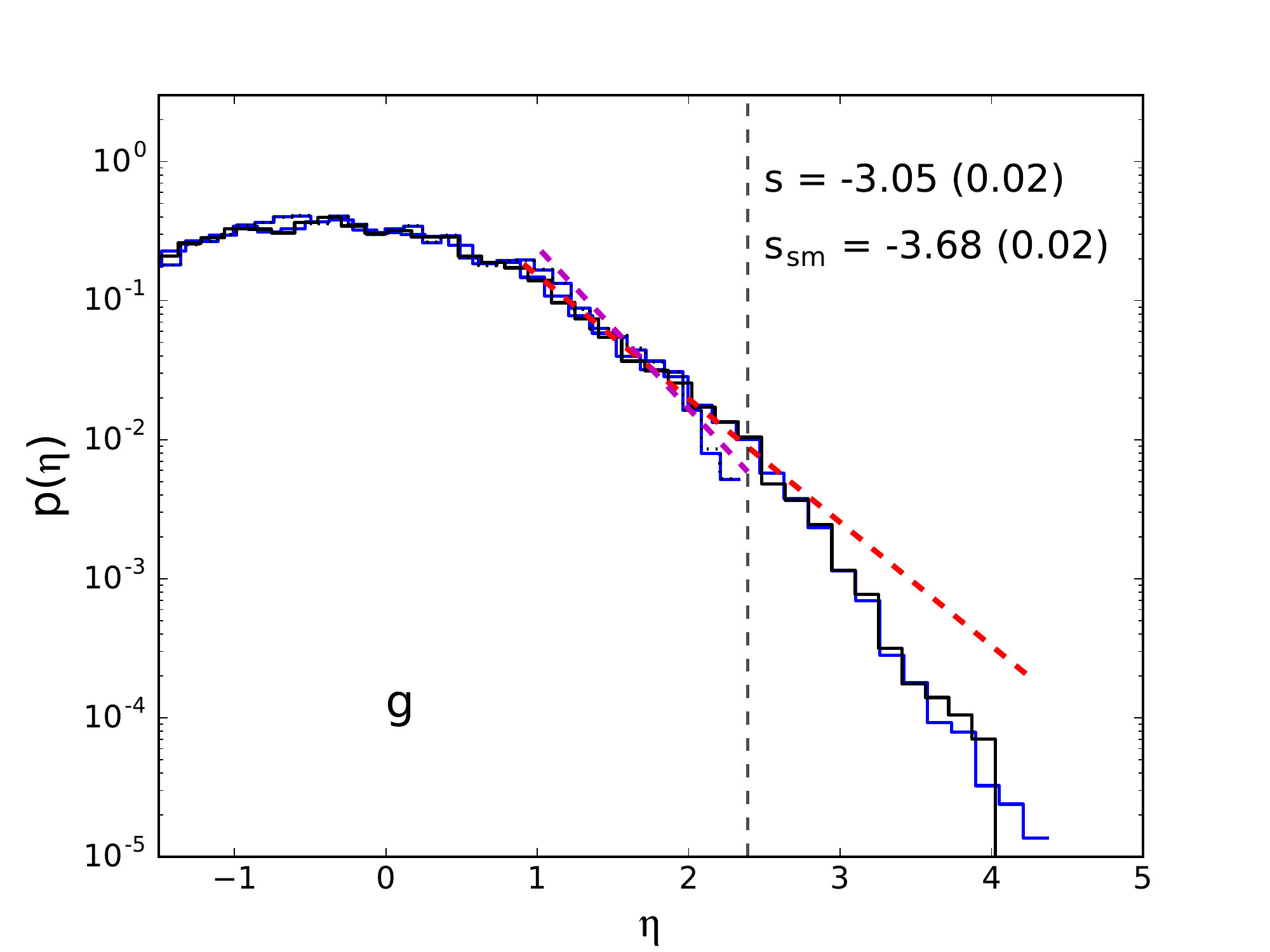} & 
\includegraphics[width=5cm]{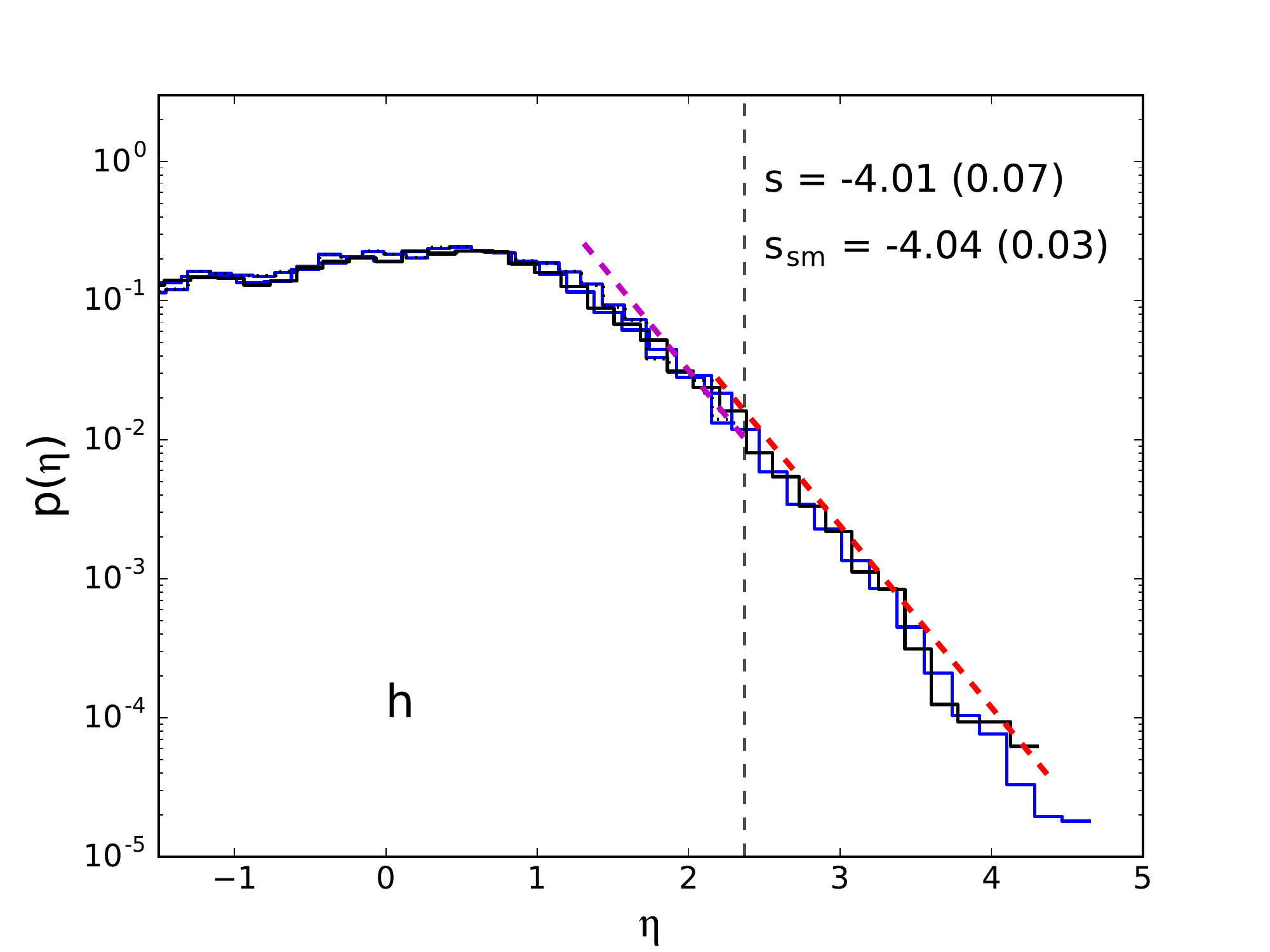}  \\
\end{tabular}
\caption{Simulated N-PDF evolution of a 10$^{4}$ $\rm M_{\odot}$ molecular cloud. The grey histogram shows the simulated $4\times 10^{6}$ data points, and the blue histogram shows the random sampled $2\times 10^{5}$  data points (comparable to our observations) to fit with power-law distributions. The N-PDFs results after smoothing are shown in dotted histograms and largely overlap with the un-smoothed results with vertical dashed lines showing the truncations. Power-law slopes for two results are indicated in the graphs, s stands for the un-smoothed result and $s_{sm}$ the smoothed results. 
}
\vspace{1.1cm}
\label{fig:simnpdf}
\end{figure*}

We tentatively suggest that the cloud integration stage and the dispersed cloud stage
stage can be separated from the stellar assembly stage and cloud pre-dispersal stage by the $s_{1}$=$-$4 power-law index of the N-PDF, and by the correlation strength at the short-lag break in the 2PT function of correlation strength 0.8.
The cloud integration stage may be further separated from the three more evolved stages by luminous-to-mass ratio $\sim$40 $\rm L_{\odot}$/$\rm M_{\odot}$.
The stellar assembly stage may have a more prominent decrease of 2PT at small scales than cloud pre-dispersal stage, with the latter possibly exhibiting a deficit of singular power-law tail of N-PDF at high column density end.
We refer to Molinari et al. (2016) for an observed relation between the luminosity-to-mass ratio and the gas temperature of high-mass molecular clumps.
The cloud pre-dispersal stage and the cloud dispersed stage may be separated by $s_{1}$$\sim$$-$4 as well.
Exactly how the cloud pre-dispersal stage and the cloud dispersed stage are separated from the N-PDF may be sensitive to the cloud initial condition and the subsequent star-formation activities.
The more precise loci in the parameter space of the power-law slope of N-PDF, the 2PT function, the N-CCDF, and the luminosity-to-mass ratios, need to be defined with a larger survey.
In this sense, the molecular cloud structures of G14.225-0.506 and G28.34+0.06 are on the transition between the {\it cloud integration} and the {\it stellar assembly} stages, when the massive molecular clumps/cores are forming.
The massive clump P2 in G28.34+0.06 already formed high-mass stars, which is a quicker process than the evolution of cloud structures.
The wiggling nature of the N-PDF of G14.225-0.506 with a ``bump'' in between $\eta$=2-3 may be due to the fact that very massive, parsec-scale forming molecular clumps have not yet reached their highest column densities (see also discussion in Section \ref{sub:sim}).
W43-South has evolved to a stage where its luminosity-to-mass ratio is higher than W43-Main, while its N-PDF is steeper, the N-CCDF is lower, and has less significant embedded massive molecular clump(s), as compared with W43-Main.
The OB cluster-forming region G10.2-0.3 appears to be already dispersed by feedback mechanisms, such that its N-PDF has a steep slope, and N-CCDF is as low as the very young source G11.11-0.12.  
We note that molecular clouds may have different morphological classes (Lin et al. 2016).
The parameters that separate the evolutionary stages may vary with the morphology classes, which needs to be examined with larger samples. 
However, the present difficulty in assembling a large sample is originated from the distance uncertainty/ambiguity.
Based on the VLA survey of $\rm NH_{3}$ lines towards 62 high-mass star-forming regions with $\rm L_{bol}\sim$10$^{4}$ $\rm L_{\odot}$, Lu et al. (2014) proposed that the clouds can be separated into the different morphological classes of {\it filaments}, {\it concentration}, and {\it dispersed} and a sub-sample not classified yet. These morphological classes may also be related to the evolutionary stages here proposed, if not purely determined by initial conditions.

\section{Conclusion}\label{section:conclusion}
We optimized the method of image combination described in  Liu et al. (2015) and Lin et al. (2016) to extend our cloud structure analysis to three IRDCs. We examined their column density distributions via N-PDF, 2PT, and N-CCDF functions, and compared to those of OB-cluster forming regions. The main findings are as follows: 
\begin{enumerate}
\item The column density probability distribution functions (N-PDFs) of G11.11-0.12 and G14.225-0.506 have steep power-law tails at their high column density end, while N-PDF of G28.34+0.06 exhibits a single power-law distribution.
Comparing to the N-PDFs of OB-cluster forming regions, the power-law tails of these IRDCs are generally steeper. The shallowing PDFs with evolutionary state indicate the dominant role of gravitational collapse on the scales we probe. Evidence of power-law tail becoming steeper again at late stage when feedback is significant is also revealed by our simulations.
\item The two-point correlation functions (2PTs) of the three IRDCs generally decrease slowly over the observed spatial scales, in stark contrast to most of the OB cluster-forming clouds in our sample, which have a steep decay of correlation strength at small spatial scales. 
This indicates that IRDCs are much less concentrated than OB cluster-forming clouds, since their column density distributions are more homogeneous at scales $>$1 pc. 
\item The column density complementary cumulative distribution function (N-CCDF) of G11.11-0.12 and G28.34+0.06 show a rapid decrease in their high column density range, similar to the most evolved OB cluster-forming region G10.2-0.3. On the other hand, G14.225-0.506 has an N-CCDF behavior closer to the other OB cluster-forming regions. 
\item Based on the observed correlations of N-PDF, 2PT, and N-CCDF measurements with luminosity-to-mass ratios, we attempt to quantitatively characterize the evolution of high-mass star-forming molecular clouds based on the evolution of their (column) density distribution function and spatial scales of gas structures.
We hypothesize four common evolutionary stages, in spite of the possibility of different initial cloud morphology and supporting mechanisms (e.g., B-field, turbulence), namely: the {\it cloud integration} stage, the {\it stellar assembly} stage, the {\it cloud pre-dispersal} stage, and the {\it dispersed-cloud} stage.  The first and fourth stages exhibit similar patterns of N-PDF, 2PT and N-CCDF, yet they can be distinguished from each other based on their rather different luminosity-to-mass ratios. The second stage has a shallower N-PDF power-law, prominent density distributions are constrained to smaller scales, as seen in their 2PT functions, and has accumulated significant gas mass in high density regimes, as shown in their N-CCDF. The third stage has a slope of N-PDF power-law still shallower than   
first stage but its 2PT has less prominent decreasing component at small scales and a possible deficit of power-law tail at high column density end as compared to second stage. Finally, power-law tail of N-PDF of final stage becomes steeper than third stage.
\end{enumerate}

The comparisons made between IRDCs and OB-cluster forming regions with these column density related measurements are necessary in an evolutionary view, in addition with comparing their different morphologies.
The physical mechanisms working at distinct evolutionary stages can be reflected by the parameters of these measurements.
We note that future observations of a larger sample would be key to span parameter space and separate the different stages unambiguously. 

\begin{acknowledgements}
\end{acknowledgements}
G.B. acknowledges the support of the Spanish Ministerio de Enconomia y Competitividad (MINECO) under grant FPDI-2013-18204. G.B. is supported by the Spanish MINECO grant AYA2014-57369-C3-1-P.
Z.Y.Z  acknowledges support from the European Research Council in the
form of the Advanced Investigator Programme, 321302, {\sc cosmicism}.
R.G.-M. acknowledges support from UNAM-PAPIIT program IA102817.
K.W. acknowledges the support from Deutsche Forschungsgemeinschaft (DFG) grant WA3628-1/1 through priority program 1573 (“Physics of the Interstellar Medium”).

\begin{appendix}
\section{Brief description of SED fitting procedure}
We briefly summarize our methods for deriving the column density and dust temperature maps here. 
We adopt a dust opacity $\kappa_{\nu}$ from Ossenkopf \& Henning (1994) where $\kappa_{230}$ = $0.9\ \rm cm^{2}g^{-1}$ is the dust opacity for thin ice mantles with $10^6$$ \rm cm^{-2}$ gas density without coagulation (frequently referred to as OH5) at 230\ GHz and a gas-to-dust ratio $M_{g}/M_{d}$ of 100.
As modified black-body assumption, the flux density $S_{\nu}$ at a certain observing frequency $\nu$ is given by
\begin{equation}
S_{\nu} = \Omega_{m} B_{\nu}(T_{d})(1-e^{-\tau_{\nu}}),
\end{equation}
and 
\begin{equation}
N_{\rm H_{2}} = \frac{\tau_{\nu}M_{g}}{\kappa_{\nu}\mu m_{\rm H}M_{d}},
\end{equation}
where
$B_{\nu}(T_{d})$ 
is the Planck function at a given temperature $T_{d}$, and
\begin{equation}
\kappa_{\nu}=\kappa_{230}(\frac{\nu}{230\ GHz})^{\beta}
\end{equation} 
\noindent where $\beta$ is the dust opacity index. We adopt a dust opacity law similar to Hildebrand (1983).
In the iterative SED fits we first use 70 $\micron$, 160 $\micron$ from PACS and combined SPIRE 350 $\micron$ with SHARC2 350 $\micron$, combined Herschel extrapolated 850 $\micron$ with SCUBA2/LABOCA and {\it{PLANCK}} 353 GHz data to simultaneously fit $\beta$, gas column density and dust temperature. 
In the final fits, we use 70 $\micron$ and combined 350 $\micron$ with derived $\beta$ from former fits to derive gas column densities and dust temperature to achieve the best angular resolution of 10$''$.
For the discussion of potential uncertainties, we refer to Section 2.4 and Appendix C in Lin et al. (2016).

\section{Morphological comparison of IRDC G14.225-0.506 and G28.34+0.06 with our simulated images}
Morphologically, our simulated images resemble with these IRDCs. We include these early epoch simulated images (direction flipped as compared to Figure. \ref{fig:simn}) side-by-side with our derived column density maps of G14.225-0.506 and G28.34+0.06. The geometrical resemblance we see here is good suggestion to support our comparisons of N-PDFs between simulated images and our observations.
\begin{figure*}
\hspace{-0.3cm}
\vspace{-0.1cm}
\begin{tabular}{p{11cm}p{11cm}}
\includegraphics[width=11cm]{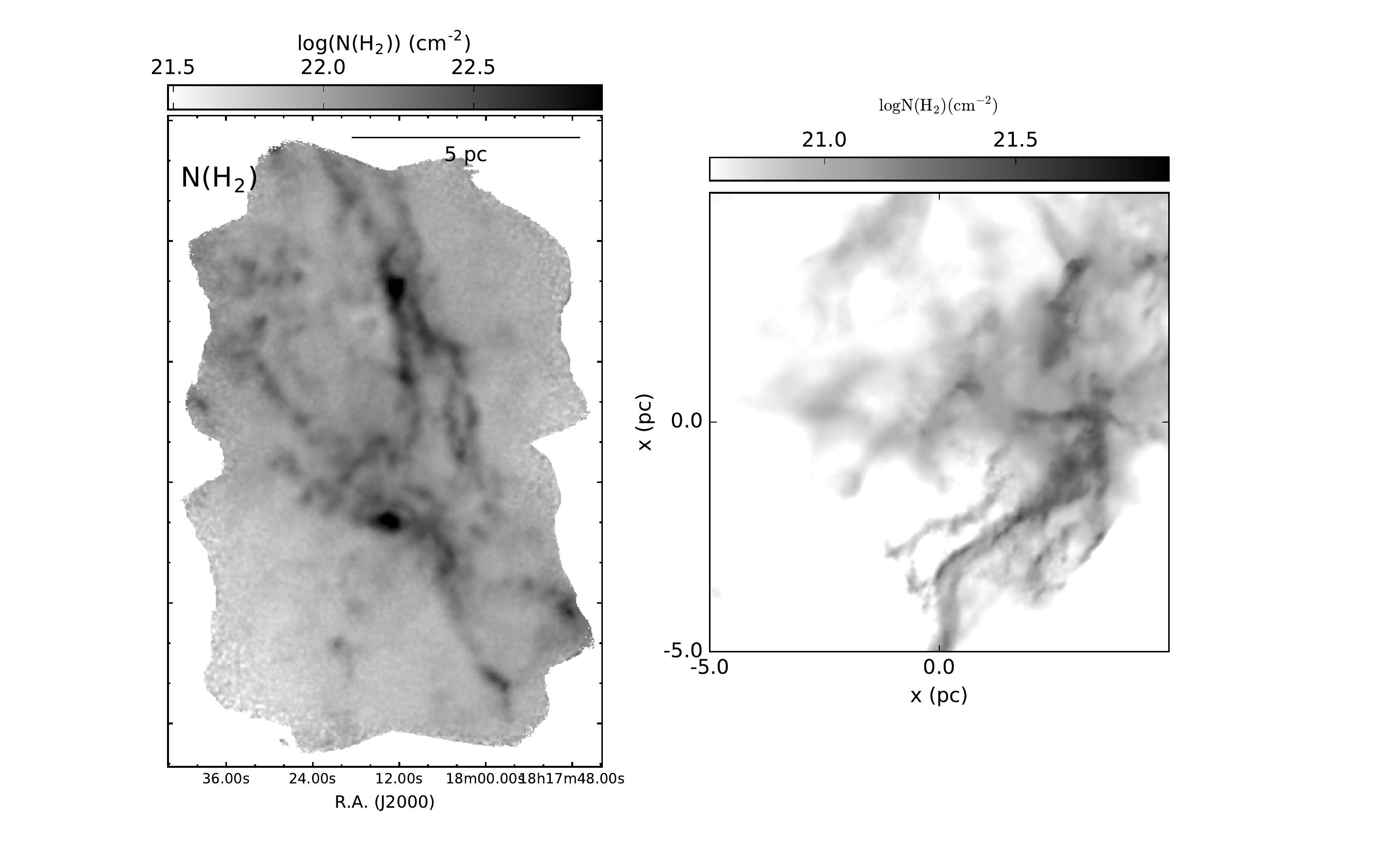}&
\hspace{-1.9cm}\includegraphics[width=11cm]{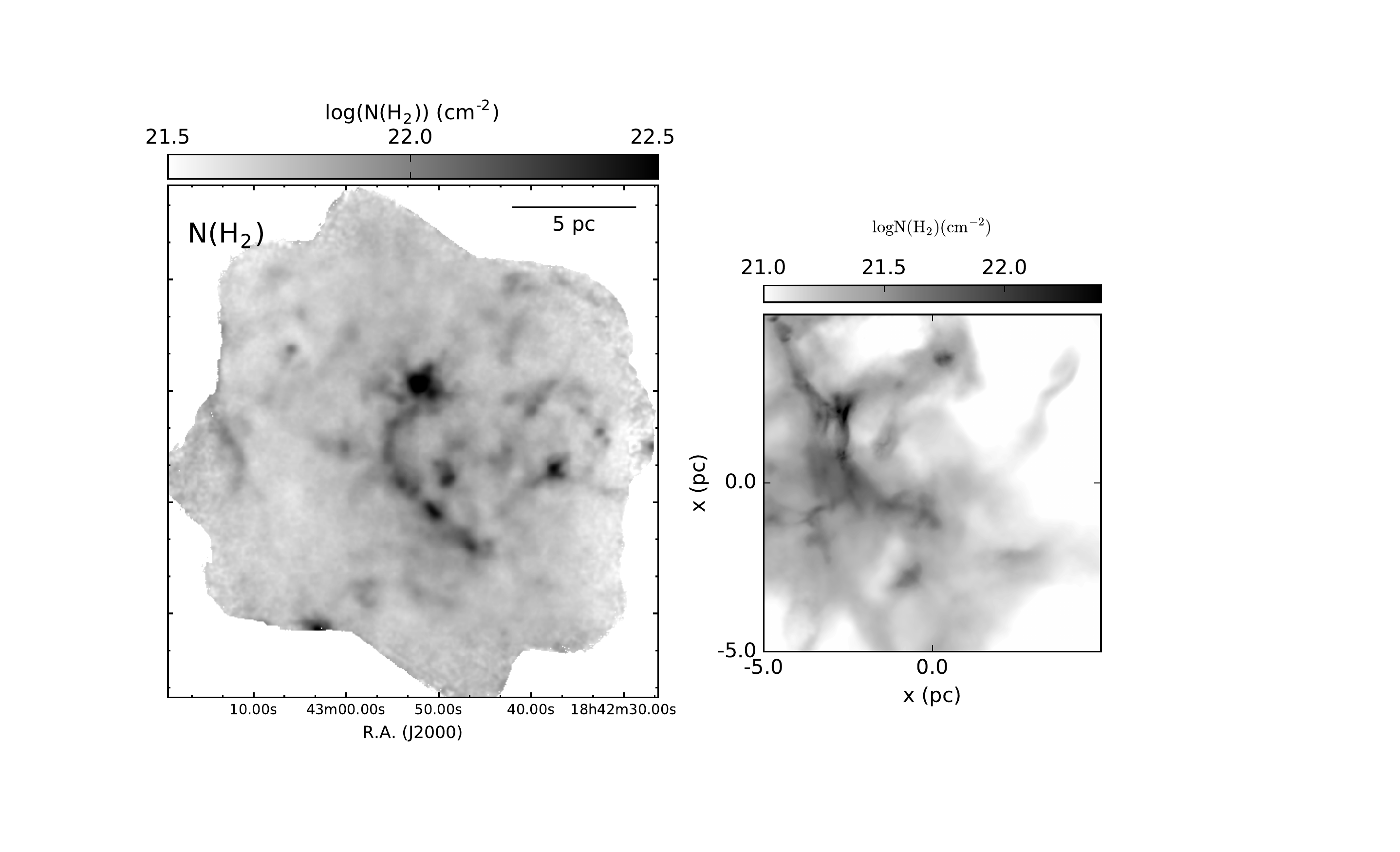}\\
\end{tabular}

\caption{{\it{Left panel}}: G14.225$-$0.506 column density map and simulated image of time epoch 3.9 Myr. {\it{Right panel}}: G28.34$+$0.06 column density map and simulated image of time epoch 2.4 Myr. Spatial scale of simulated images are adjusted to be the same with corresponding sources.}
 
\end{figure*}

 \end{appendix}

\end{document}